\documentclass[reprint,superscriptaddress,aps,prb,floatfix]{revtex4-2}

\usepackage{mathtools}
\usepackage{graphicx}
\usepackage{hyperref}
\usepackage[export]{adjustbox}
\usepackage{tikz}
\usepackage{braket}
\usepackage{textcomp,gensymb}
\usepackage{dsfont}
\usepackage{algpseudocode}
\usepackage{bbold}

\makeatletter
\def\parsecomma#1,#2\endparsecomma{\def\page@x{#1}\def\page@y{#2}}
\tikzdeclarecoordinatesystem{page}{
    \parsecomma#1\endparsecomma
    \pgfpointanchor{current page}{north east}
    \pgf@xc=\pgf@x%
    \pgf@yc=\pgf@y%
    \pgfpointanchor{current page}{south west}
    \pgf@xb=\pgf@x%
    \pgf@yb=\pgf@y%
    \pgfmathparse{(\pgf@xc-\pgf@xb)/2.*\page@x+(\pgf@xc+\pgf@xb)/2.}
    \expandafter\pgf@x\expandafter=\pgfmathresult pt
    \pgfmathparse{(\pgf@yc-\pgf@yb)/2.*\page@y+(\pgf@yc+\pgf@yb)/2.}
    \expandafter\pgf@y\expandafter=\pgfmathresult pt
}
\makeatother

\newcommand{\bvec}[1]{\boldsymbol #1}
\DeclareMathOperator{\hadamard}{\circ}
\renewcommand{\paragraph}[2]{{\bf#2}}

\begin{document}

\title{Unconventional Superconductivity in Magic-Angle Twisted Trilayer Graphene}

\author{Ammon Fischer}
\affiliation{Institute for Theory of Statistical Physics, RWTH Aachen
University, and JARA Fundamentals of Future Information Technology, 52062
Aachen, Germany}
\author{Zachary A. H. Goodwin}
\affiliation{Departments of Materials and Physics and the Thomas Young Centre
for Theory and Simulation of Materials, Imperial College London, South
Kensington Campus, London SW7 2AZ, UK}
\author{Arash A. Mostofi}
\affiliation{Departments of Materials and Physics and the Thomas Young Centre
for Theory and Simulation of Materials, Imperial College London, South
Kensington Campus, London SW7 2AZ, UK}
\author{Johannes Lischner}
\affiliation{Departments of Materials and Physics and the Thomas Young Centre
for Theory and Simulation of Materials, Imperial College London, South
Kensington Campus, London SW7 2AZ, UK}
\author{Dante M. Kennes}
\altaffiliation{Corresponding author:
\href{mailto:dante.kennes@rwth-aachen.de}{dante.kennes@rwth-aachen.de}}
\affiliation{Institute for Theory of Statistical Physics, RWTH Aachen
University, and JARA Fundamentals of Future Information Technology, 52062
Aachen, Germany}
\affiliation{Max Planck Institute for the Structure and Dynamics of Matter,
Center for Free Electron Laser Science, 22761 Hamburg, Germany}
\author{Lennart Klebl}
\altaffiliation{Corresponding author:
\href{mailto:klebl@physik.rwth-aachen.de}{klebl@physik.rwth-aachen.de}}
\affiliation{Institute for Theory of Statistical Physics, RWTH Aachen
University, and JARA Fundamentals of Future Information Technology, 52062
Aachen, Germany}

\date{\today}

\begin{abstract}
  Magic-angle twisted trilayer graphene (MATTG) recently emerged as a highly
  tunable platform for studying correlated phases of matter, such as correlated
  insulators and superconductivity.  Superconductivity occurs in a range of
  doping levels that is bounded by van Hove singularities which stimulates the
  debate of the origin and nature of superconductivity in this material.  In
  this work, we discuss the role of spin-fluctuations arising from atomic-scale
  correlations in MATTG for the superconducting state.  We show that in a phase
  diagram as function of doping ($\nu$) and temperature, nematic superconducting
  regions are surrounded by ferromagnetic states and that a superconducting dome
  with $T_c \approx 2\,\mathrm{K}$ appears between the integer fillings $\nu
  =-2$ and $\nu = -3$.  Applying a perpendicular electric field enhances
  superconductivity on the electron-doped side which we relate to changes in the
  spin-fluctuation spectrum. We show that the nematic unconventional
  superconductivity leads to pronounced signatures in the local density of
  states detectable by scanning tunneling spectroscopy measurements.
\end{abstract}

\maketitle

\section*{Introduction}
Since the discovery of superconductivity and correlated insulating states in
magic-angle twisted bilayer graphene
(MATBG)~\cite{cao2018unconventional,cao2018mott}, twisted van der Waals
materials have become indispensable for the design of novel quantum materials at
will~\cite{moiresim}. In the quickly developing field of
twistronics~\cite{Carr2017twistronics}, tremendous
theoretical~\cite{Guinea2018,Kennes2018,fischer2021spin,klebl2020importance,
klebl2019inherited,Gonzalez2017,Xie2020HF,Stauber2019KL,Zhang2020,
Goodwin2019attractive,Stauber2020,Cea2020,klebl2020frg,Yan2018,Wu2018ph,
Choi2018ph,Padhi2018WC,yu2021nematicity,samajdar2021electricfieldtunable,
qin2021critical,lin2019chiral,lin2020parquet} and
experimental~\cite{lu2019superconductors,Cao2020strange,Polshyn2019,
yankowitz2019tuning,liu2021tuning,stepanov2020untying,saito2020independent,
Das2020Chern,Wu2020Chern,Nuckolls2020Chern,arora2020superconductivity,
Serlin2020,Sharpe2019ferro,Zondiner2020,Wong2020,Xie2019spectrosopic,
Kerelsky2019maximized,Jiang2019charge,Choi2019correlations,cao2020nematicity}
efforts have been undertaken to unravel the nature of strong
correlations~\cite{Carr2020NatRev,Balents2020rev} and to access new moir\'e
engineered structures with twisted double-bilayer
graphene~\cite{burg2019correlated, shen2020correlated, liu2020tunable,
rubioverdu2020universal}, twisted trilayer
graphene~\cite{Park2021,hao2021engineering,cao2021large, tsai2019correlated,
shi2020tunable}, transition metal dichalcogenide homo- and
heterobilayers~\cite{wang2020correlated, tang2020, xian20, regan2020,Vitale2021}
as well as other materials~\cite{Xian18, Ni2019, kennes19} at the frontier of
condensed matter research~\cite{moiresim}.

These systems are fascinating because of the precise control of electronic
properties and correlations that can be achieved by tuning twist
angle~\cite{cao2018unconventional,cao2018mott}, doping
level~\cite{cao2018unconventional,cao2018mott,lu2019superconductors},
temperature~\cite{Cao2020strange,Polshyn2019},
pressure~\cite{yankowitz2019tuning,Carr2018Pressure} and external
screening~\cite{Goodwin2020screening,liu2021tuning,stepanov2020untying,
saito2020independent}.  The appearance of almost flat bands at so-called ``magic
angles", first predicted in early theoretical
works~\cite{bistritzer2011moire,dos2007graphene, shallcross2010electronic,
morell2010flat}, puts a variety of exotic correlated phases within experimental
reach, including correlated
insulators~\cite{lu2019superconductors,Cao2020strange,Polshyn2019,
yankowitz2019tuning,liu2021tuning,stepanov2020untying,
arora2020superconductivity,
Zondiner2020,Wong2020,Xie2019spectrosopic,Kerelsky2019maximized,Jiang2019charge,
Choi2019correlations,cao2020nematicity},
orbital ferromagnetism~\cite{Sharpe2019ferro, Serlin2020, lu2019superconductors,
Kang2019strong} and magnetic field induced Chern
insulators~\cite{Das2020Chern,Wu2020Chern,Nuckolls2020Chern, park2020flavour,
saito2020independent}.

\begin{figure*}[ht!]
  \includegraphics[width=\textwidth]{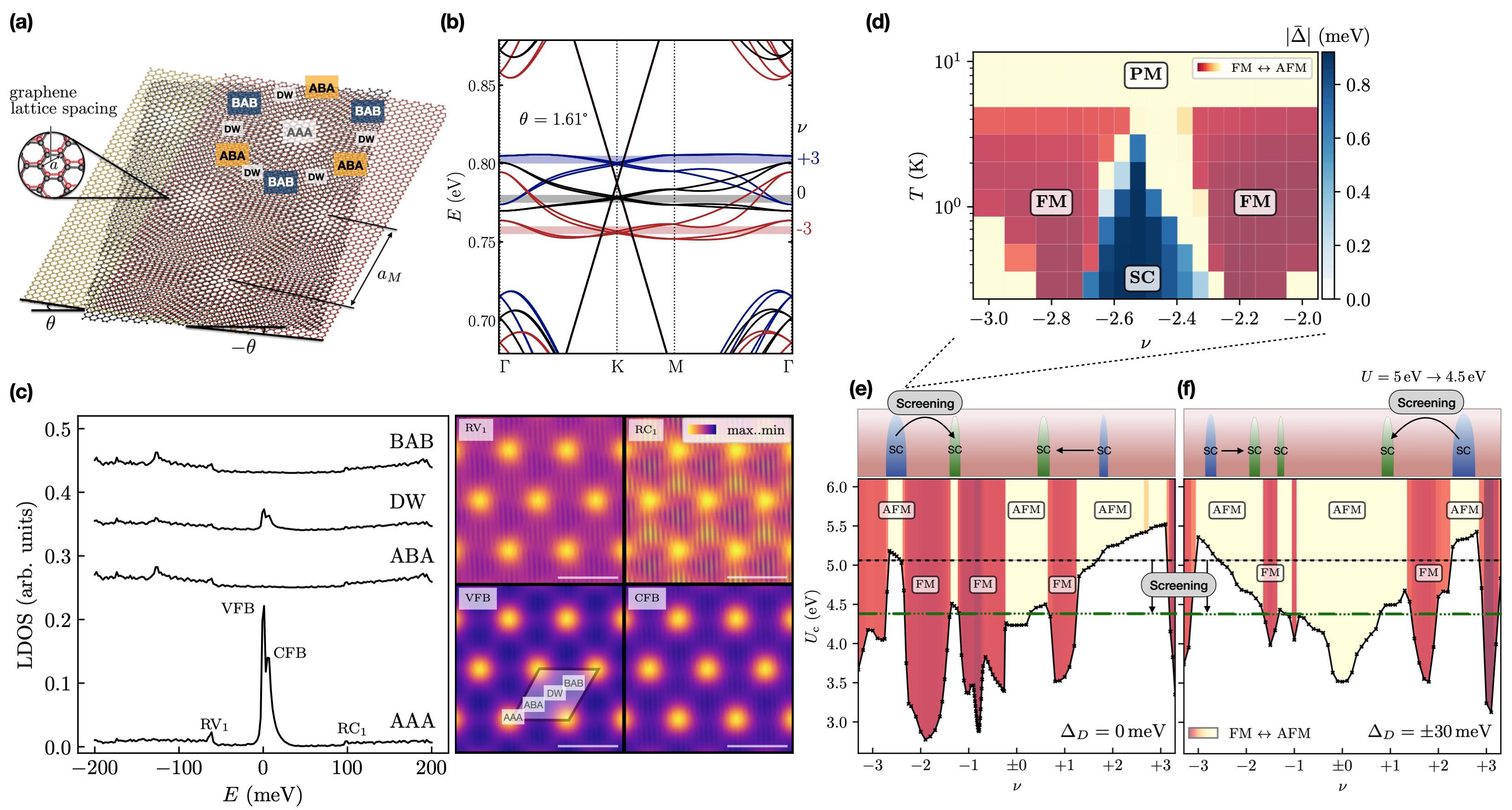}
    \caption{{\bf Unconventional superconductivity in magic-angle twisted
    trilayer graphene (MATTG). (a)} The atomic structure of MATTG consists of
    three superimposed graphene sheets. While the outer graphene sheets are
    aligned (AA stacked) and untwisted, the inner layer (black) is twisted at an
    angle $\theta$ ($-\theta$) relative to the lower (upper) sheet. {\bf (b)}
    Band structure of MATTG at $\theta=1.61\degree$. Long-range
    electron-electron interactions are taken into account through Hartree
    corrections that lead to a doping-dependent band structure. The flat band
    dispersion is strongly affected by the filling factor $\nu$ (blue: $\nu=+3$,
    black: $\nu=0$, red: $\nu=-3$), whereas the Dirac cone remains unaffected.
    {\bf (c)} Local density of states (LDOS) in the outer layer of MATTG for
    $\nu=-2.5$. The left panel clearly shows that the flat bands are mostly
    localized in the AAA regions. The right panel shows the spatial distribution
    of the LDOS in the top layer at energies of the remote valence band
    ($\text{RV}_1$), remote conduction band ($\text{RC}_1$), valence flat band
    (VFB) and conduction flat band (CFB), reflecting the $C_{3z}$ symmetry of
    the non-interacting Hamiltonian.  {\bf (d)} Phase diagram of MATTG for
    $\theta=1.61\degree$ around $\nu=-2.5$. Our numerical calculations reveal a
    superconducting dome driven by low-energy AFM spin-fluctuations that ranges
    from  $\nu=-2.35$ to $\nu=-2.7$ with an upper critical temperature of
    $T_\mathrm{c}\approx2\,\mathrm{K}$. The average amplitude of the order
    parameter $|\bar{\Delta}|$ is reduced with increasing temperature and
    vanishes towards the integer fillings $\nu=-3$ and $\nu = -2$, where
    ferromagnetic phases (FM) dominate. At high temperatures, the system remains
    paramagnetic (PM). {\bf (e,f)} Magnetic correlations in MATTG at zero (left)
    and nonzero $\Delta_D=\pm30\,\mathrm{meV}$ (right) perpendicular
    displacement field. The lower panel displays the critical Hubbard
    interaction strength $U_\mathrm{c}$ needed for the onset of magnetic order
    as a function of filling.  The type of magnetic order is color-coded through
    an order parameter that continuously interpolates between AFM and FM order
    (red: FM, yellow: AFM, for a definition see Supplemental Methods). The
    upper panels display a sketch [which is confirmed quantitatively for filling
    between $\nu=-3$ and $\nu = -2$ in {\bf (d)}] of parameter regions that can
    host unconventional SC driven by AFM spin-fluctuations for
    $U\approx5\,\mathrm{eV}$ (dashed black horizontal line). As magnetic
    interactions can only provide the pairing glue for superconductivity as long
    as the system remains paramagnetic, i.e. $U<U_\mathrm{c}$, this mechanism
    supports SC at $\nu=-2-\delta$ and around $\nu=+2+\delta$.  Screening
    effects caused by the dielectric environment may change the position of the
    superconducting domes to different fillings (blue to green SC domes).
    Applying an electric field to the sample (panel {\bf (f)}) enhances
    superconducting regions at the electron-doped side at $\nu=+2+\delta$.}
    \label{fig:sttg}
\end{figure*}

Among the findings that have sparked the most interest in the field of
twistronics is the discovery of robust and reproducible superconductivity in
MATBG~\cite{cao2018unconventional,yankowitz2019tuning, lu2019superconductors},
with preliminary evidence for possible superconductivity also present in
twisted-double bilayer graphene~\cite{burg2019correlated, shen2020correlated,
liu2020tunable}, ABC trilayer graphene aligned to hexagonal boron
nitride~\cite{Chen2019ABC} and twisted transition metal
dichalcogenides~\cite{wang2020correlated}. Very recently, another graphitic
moir\'e system that features reproducible, highly tunable superconductivity (as
well as correlated insulators) has been discovered: magic-angle twisted trilayer
graphene (MATTG)~\cite{Park2021,hao2021engineering,cao2021large}, where the
twist angle alternates by $+\theta$ and $-\theta$ between each graphene layer
[see Fig.~\ref{fig:sttg}~{\bf(a)}].  Experiments on MATTG find superconductivity
at doping levels between integer fillings of  $\nu = -2 $ and  $\nu = -3$, where
$\nu$ corresponds to the number of electrons per moir\'e unit cell relative to
charge neutrality~\cite{Park2021,hao2021engineering, cao2021large}.
Additionally, Refs.~\onlinecite{Park2021,cao2021large} report superconductivity
between fillings of $\nu = 2$ and $\nu = 3$, which is much weaker in
Ref.~\onlinecite{hao2021engineering} but can be stabilized upon application of a
perpendicular displacement field.  Furthermore, displacement fields are found to
give rise to additional superconducting features between fillings of $\nu = 1$
and $\nu =2$ and between $\nu = -1$ and $\nu = -2$~\cite{Park2021}.  This
demonstrates that (i) superconductivity in MATTG seems to preferentially appear
in between integer fillings of the flat bands and that (ii) superconductivity in
MATTG can be readily tuned through a perpendicular displacement field, which
makes MATTG a particularly attractive platform for studying strongly correlated
physics. 

The coexistence of correlated insulating and superconducting states in MATTG has
further elicited questions about their intrinsic relationship in graphene-based
moir\'e materials. In particular, the question of whether the superconducting
states are of unconventional nature and driven by electron-electron interaction,
or conventional and mediated by electron-phonon coupling, is still intensely
debated at the moment even for MATBG~\cite{Balents2020rev, stepanov2020untying,
qin2021critical}. 

The aim of this work is to shed light on this controversial question by
presenting how unconventional superconductivity in MATTG can arise from spin
fluctuation exchange on the atomic scale. Starting from a fully atomistic
tight-binding description of the system, we investigate the effect of
long-ranged electron-electron interactions on the phase diagram of MATTG. To
this end, we derive a microscopic pairing interaction $\hat \Gamma_2$ in the
fluctuation exchange approximation (FLEX) and solve the non-linear Bogoliubov-de
Gennes equations as function of filling and temperature. Our results indicate
that spin-singlet superconductivity can be driven by magnetic fluctuations in
between integer fillings. At the same time, superconductivity is shown to depend
sensitively on the value of the carbon $\mathrm{p}_z$ Hubbard-$U$, which is
influenced by experimental details such as environmental screening and the
application of fields, and that it can be moved from one between-integer filling
to another flexibly. For a suitably chosen interaction strength and without a
displacement field, we find superconductivity is strongest in between fillings
of $\nu=-2$ and $\nu=-3$ (with critical temperature $T_c\approx2\,\mathrm{K}$)
as well as around $\nu = +2+\delta$, while being surrounded by ferromagnetic,
semi-metallic or metallic states. We show that, for fixed interaction strength,
a perpendicular electric field weakens the superconducting dome with on the
hole-doped side, but enhances superconductivity at doping levels around $\nu=+2
+ \delta$.  Hereby we demonstrate that correlated and superconducting features
driven by electronic interactions in MATTG are highly tunable by a perpendicular
electric field which corroborates recent experimental findings~\cite{Park2021,
hao2021engineering, cao2021large}, in particular those of
Ref.~\onlinecite{hao2021engineering}. We then analyze the nature of the
superconducting state further and show that according to our atomistic
calculations MATTG hosts unconventional, nematic $d$-wave superconductivity that
displays clear signatures of $C_{3z}$ symmetry breaking in the local density of
states.

\section*{Results}
\paragraph*{Atomic, Electronic and Magnetic Structure.}--- 
The atomic structure of MATTG is constructed from three stacked sheets of
graphene, where the outer layers ($l=1,3$) are perfectly aligned and the middle
layer ($l=2$) is twisted by $\pm \theta$ relative to the encapsulating layers,
as schematically depicted in Fig.~\ref{fig:sttg}~{\bf(a)}. Different regions in
the moir\'{e} unit cell of MATTG can be labeled according to the stacking
sequence of the trilayer. At the center of rotation, the stacking is of AAA
type, where carbon atoms of all layers reside vertically displaced on top of
each other, as shown in the inset of Fig.~\ref{fig:sttg}~{\bf(a)}.  In between
the AAA regions, the local geometry successively changes from ABA to domain wall
(DW) and BAB stacking.  The atomic structure described is invariant under the
mirror reflection symmetry $\sigma_h$ that exchanges the upper and lower layer
($1 \leftrightarrow 3, 2 \leftrightarrow 2$) and under three-fold rotations
around the $\hat z$ axis described by the point group $C_{3z}$. 

To account for lattice reconstruction effects, which have been shown to be
important in the field of twistronics~\cite{Carr2020NatRev}, we first relax the
positions of MATTG using classical force fields (see Methods). According to the
reflection symmetry $\sigma_h$, we find that the central layer remains flat,
while the atoms of the outer layers undergo significant out-of-plane
displacements, which is in good agreement with other
work~\cite{Kruchkov2020,lopez2020electrical}. In fact, recent
results~\cite{Kruchkov2020} indicates that perfect alignment between the outer
layers, as considered in this study, is energetically favoured over a relative
shift between the outer layers, i.e. AAB/BAA stacking, and hence should be the
most relevant stacking configuration that is naturally realized in experimental
samples.

To model the electronic structure of MATTG we use a microscopic tight-binding
(TB) parametrization for the $\mathrm{p}_z$ orbitals of the carbon
atoms~\cite{trambly2010localization} combined with \textit{ab initio} density
functional theory (DFT) simulations (see Supplementary Methods and Methods).
To match the dispersion obtained from DFT, we complement the TB model with an
additional onsite potential of $-35\,\mathrm{meV}$ acting on the middle layer.
As a result the flat bands intersect with the Dirac cone below the Dirac point,
which was also noted in previous work for MATTG by Lopez-Bezanilla and
Lado~\cite{lopez2020electrical}. As shown in Fig.~\ref{fig:sttg}~{\bf(b)}, we
find that for $\theta = 1.61\degree$, the low-energy electronic structure of
MATTG consists of a set of flat bands (similar to MATBG), which are intersected
by a Dirac cone with a large Fermi
velocity~\cite{Kruchkov2020,zhu2020twisted,lopez2020electrical} compared to the
flat band kinetic energy scales. The twist angle investigated here is very close
to the magic angle of $1.54\degree$ which exhibits the smallest flat band
width~\cite{khalaf2019magic}. 

In MATBG, long-ranged electron-electron interactions were found to strongly
alter the electronic structure, which has important consequences for the
observation of broken symmetry phases. MATTG is an ostensibly similar system,
and therefore, we also treat the long-ranged part of the electron interactions
using a self-consistent Hartree theory in the atomistic tight-binding framework
(see Supplementary Methods and Methods)~\cite{Guinea2018,
goodwin2020hartree}. Through the Hartree potential, the electronic structure of
MATTG in the normal state acquires a filling dependence, which shifts the
electronic bands to higher or lower energies depending on the doping level. We
show in Fig.~\ref{fig:sttg}~{\bf(b)} that the dispersion of the flat bands of
MATTG in the normal state is indeed very sensitive to long-ranged electron
interactions: removing electrons lowers the K-point energies of the flat bands
relative to the $\Gamma$-point, which is similar to the doping dependence of
MATBG~\cite{Guinea2018, goodwin2020hartree}, and is accompanied by a global
shift of the flat bands relative to the Dirac cone to more negative energies. In
contrast, adding electrons increases the K-point energies of the flat bands
relative to the $\Gamma$-point, and shifts the whole flat band manifold to
higher energies relative to the Dirac cone. The Dirac cone with its large Fermi
velocity is insensitive to long-ranged electron interactions.

The local density of states in the outer layer of MATTG, as shown in
Fig.~\ref{fig:sttg}~{\bf(c)} for a doping level of $\nu=-2.5$, exhibits large
peaks in the AAA regions for energies within the range of the flat bands (the
conduction flat band CFB and valence flat band VFB). This behaviour is
consistent with the earlier prediction that the flat-band physics in MATTG is
similar to that of MATBG~\cite{khalaf2019magic}. Although the flat bands have
their largest spectral weight in the central layer, significant weight is
distributed on the outer layers such that the van Hove singularities associated
with the flat bands should be observable in scanning tunneling microscopy (STM)
experiments.  This is shown in the left panel of Fig.~\ref{fig:sttg}~{\bf(c)},
where the flat bands give rise to $C_{3z}$ symmetric signatures in the LDOS. At
larger energies (remote conduction band RC$_1$ and remote valence band RV$_1$),
the states become more delocalized [see right panel of
Fig.~\ref{fig:sttg}~{\bf(c)}].

Having captured the effect of long-range electron-electron interactions in the
doping-dependent band structure, we study the influence of the remaining
short-ranged terms, i.e., a repulsive Hubbard-$U$, by calculating the atomistic
susceptibility $\hat{\chi}_0$ in the magnetic channel. To this end, we employ
the random phase approximation (RPA) in its static, long-wavelength
limit~\cite{klebl2019inherited,klebl2020importance} (see Methods).  As the
effect of the Hartree potential on the band structure of MATTG is very similar
to the MATBG case, the consequence we expect is a broader range of twist angles
showing correlated states driven by short-range
interactions~\cite{klebl2020importance}.  For given filling and temperature
$(\nu, T)$, the magnetic susceptibility $\hat{\chi}_0$ contains information
about (i) the critical interaction strength $U_\mathrm{c}$, that is the minimal
value of $U$ needed to drive the system from the paramagnetic regime
($U<U_\mathrm{c}$) into magnetic order ($U \geq U_\mathrm{c}$) and (ii) the type
of magnetic order depending on the distribution of magnetic moments in the
moir\'e unit cell.

In Figure~\ref{fig:sttg}~{\bf(e)} we show $U_\mathrm{c}$ as a function of
filling at $T=1.3\,\mathrm{K}$. The color maps indicate the type of magnetic
ordering [yellow: antiferromagnetic  (AFM) and red: ferromagnetic (FM)].  A
precise definition of the order parameter that continuously interpolates from FM
to AFM order is given in the Supplemental Methods. We find that small values of
$U_\mathrm{c}$ are observed at (or close to) the integer fillings $\nu = \pm
3,-2,\pm 1$ driving FM order. Small values of $U_\mathrm{c}$ indicate that the
system is very susceptible to this kind of magnetic order as already a small
interaction value $U$ is sufficient to trigger the magnetic instability.
Interestingly, for $\nu = -3,-2,\pm 1$ these dips in $U_\mathrm{c}$ are
surrounded by AFM regions exhibiting a much larger $U_\mathrm{c}$. Such
behaviour was previously observed in MATBG~\cite{fischer2021spin} and indicates
that, depending on the value of the Hubbard-$U$, these AFM instabilities may not
be strong enough to actually occur, which opens the door for possible
spin-fluctuation mediated superconductivity.

To investigate the influence of an external, perpendicular electric field, we
set the onsite energies of the outer layers to $\Delta_D=\pm30\,\mathrm{meV}$,
which models the presence of a displacement field similar in magnitude (see
Supplementary Methods) to that applied in the experiments of
Ref.~\citenum{hao2021engineering} to achieve their highest superconducting
transition temperatures. In Figure~\ref{fig:sttg}~{\bf(f)} we show the same RPA
analysis for this additional interlayer potential. In contrast to what was found
without an electric field, we now observe small values of $U_\mathrm{c}$ close
to $\nu = +2$ electrons.  We observe an increase in the small value of
$U_\mathrm{c}$ at $\nu=+1$, and the value at $\nu=+3$ remains relatively
unchanged by the displacement field.  Moreover, on the hole-doped side
$U_\mathrm{c}$ is increased strongly almost over the whole doping range, which
is where the lowest values of $U_\mathrm{c}$ were observed without a
displacement field. Interestingly, now an AFM region which is surrounded by FM
regions emerges around $\nu=+2.5$. Within this AFM region, $U_\mathrm{c}$ is
large, which yields a paramagnetic phase with AFM fluctuations for
$U<U_\mathrm{c}$.

\paragraph*{Spin-fluctuation induced Superconductivity.}--- 
To model unconventional superconductivity in MATTG, we assume that Cooper pairs
are formed due to intricate effects arising from microscopic electron-electron
interactions. Hence, this approach differs from the conventional BCS theory with
electron-phonon coupling as the predominant pairing mechanism.  In the vicinity
of magnetic (semi-metallic or metallic) instabilities described above, a
spin-polarized electron may travel through the graphene lattice and polarize the
other electrons' spin around it such that an attractive interaction arises
between the electrons and they finally form a bound state \cite{Berk1966}.
Technically, we capture this effect of transverse and longitudinal
spin-fluctuations by the microscopic pairing vertex $\hat{\Gamma}_2$, which we
derive using the the fluctuation-exchange approximation (FLEX). We then solve
the non-linear Bogoliubov de-Gennes (BdG) equations self-consistently, using the
full atomistic FLEX pairing vertex to obtain information about the nature of the
superconducting state (for details see Method section and Supplementary
Methods).

When MATTG is doped near $\nu= - 2.5$ in the absence of a displacement field,
our unconventional BCS theory reveals nematic superconductivity that originates
from AFM spin-fluctuation exchange. As shown in Fig.~\ref{fig:sttg}~{\bf(d)}, as
a function of filling and temperature, the system features a superconducting
dome enclosed by FM instabilities at integer fillings $\nu = -2$ and $\nu=-3$.
The transition temperature of the superconducting phase is substantially
influenced by the spin-fluctuation spectrum. As AFM tendencies are weakened with
increasing temperature and FM instabilities dominate around the integer
fillings, the superconducting region is effectively confined between $\nu = -2$
and $\nu=-3$ with an upper critical temperature of $T_c \approx 2\,\mathrm{K}$.
For even larger temperatures, the flat bands of the system are no longer
resolved due to temperature broadening and the system continuously returns to
ordering tendencies inherited from the untwisted
system~\cite{klebl2019inherited}. This can be clearly recognized by the FM
ordering tendencies disappearing at high $T$.

The driving force behind this superconducting phase originates from low-energy
AFM spin-fluctuations~\cite{Berk1966} that can provide an attractive potential
for electrons in MATTG. The non-uniform real-space profile of the spin-mediated
pairing vertex $\hat{\Gamma}_2$ in the moir\'e unit cell (see Supplementary
Methods) shows that these attractive components are strongest on
nearest-neighbor bonds in the single graphene sheets of MATTG, thus suggesting
in-plane Cooper pairs. The low-energy spin-fluctuations are strengthened in the
vicinity of the magnetic instability, i.e., when the value of the repulsive
Hubbard-$U$ is slightly below the critical interaction strength $U\lesssim
U_{\mathrm{c}}$.  In this parameter regime, the system shows no magnetic order,
but the attractive interactions mediated by the spin fluctuations can provide
the pairing glue for unconventional spin-singlet superconductivity between the
integer fillings.  At the same time, singlet Cooper pairs may not be relevant
close to the FM instabilities as the effective interaction $\hat{\Gamma}_2$ is
purely repulsive in the singlet channel.

\begin{figure*}[ht!]
  \includegraphics[width=\textwidth]{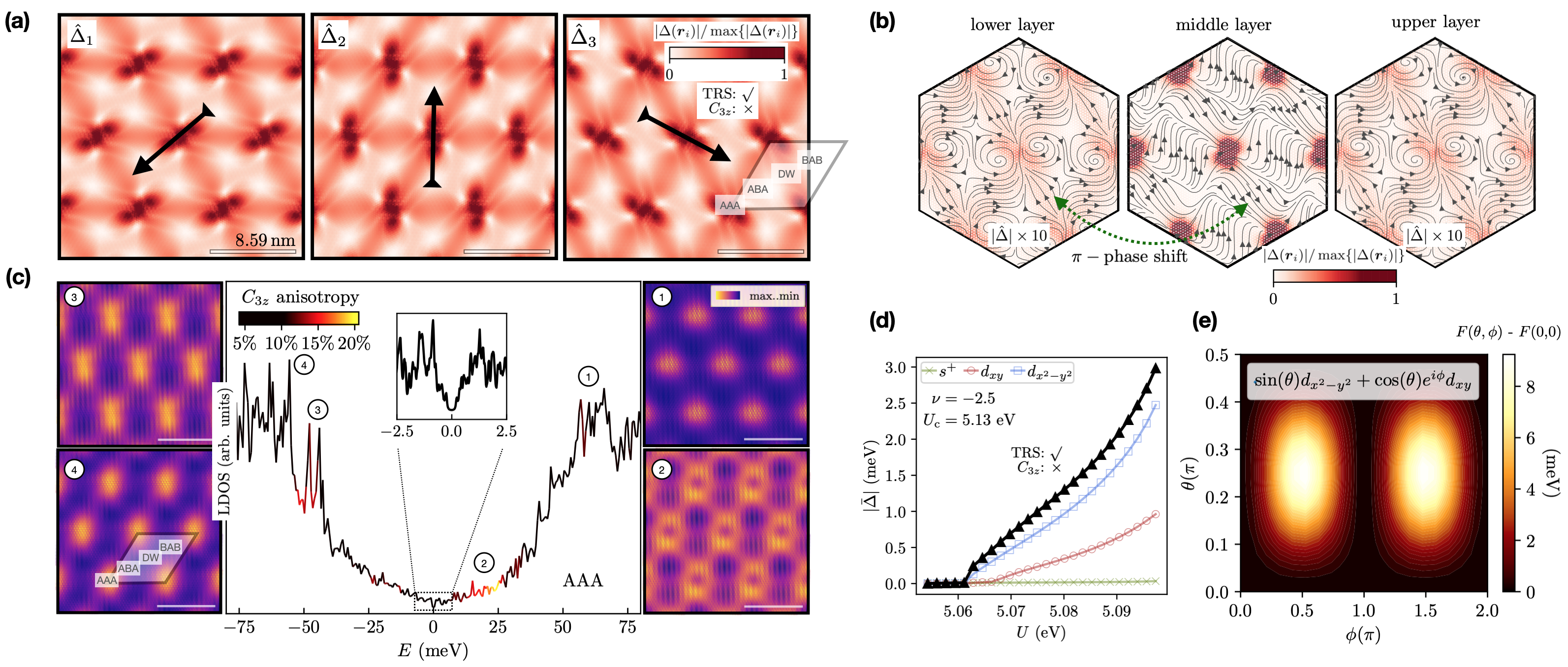}
  \caption{{\bf{}Properties of the superconducting order parameter in MATTG} for
  ${\nu=-2.5}$ and $T=0.2\,\mathrm{K}$. {\bf (a)} Spatially resolved atomistic
  gap $|\Delta(\bvec{r_i})|/\max\{|\Delta(\bvec{r_i})|\}$ in the outer layer of
  MATTG. Our analysis reveals three degenerate real-valued ground states that
  break $C_{3z}$ rotational symmetry on the moir\'{e} scale along the nematic
  axis $C_2$ (black arrow).  {\bf{}(b)} Layer-wise representation of the order
  parameter field. The value of $|\Delta(\bvec{r_i})|$ is larger by a factor of
  $\sim10$ in the middle layer compared to the outer layers with most weight
  being concentrated in the AAA regions. The phase of the superconducting gap is
  shifted by $\pi$ between single layers. Additionally to the nematicity on the
  moir\'e scale, we find strong local atomic-scale nematicity in the orientation
  of the $d$-wave components $\bvec{\tau}(\bvec r_i)$ (black streamlines). The
  moir\'e nematicity is clearly visible due to the emergence of a
  vortex-antivortex pattern near the AAA regions.  {\bf{}(c)} Local density of
  states (LDOS) in the superconducting state $\hat\Delta_3$ from panel~{\bf(a)}
  in the AAA region of the outer layer of MATTG. Peaks at selected energies
  (1)-(4), which correspond to flat bands being gapped by the order parameter,
  show clear signatures of $C_{3z}$ symmetry breaking. In subpanel (2) the
  influence of the $C_2$ nematic axis is visible: For low energies,
  quasiparticles are forbidden to occupy the AAA regions along the direction of
  $C_2$ due to the presence of the superconducting condensate, but accumulate on
  both sides where the gap amplitude vanishes. This leads to a strong degree of
  $C_{3z}$ symmetry breaking as visualized by the color code in the main panel.
  {\bf{}(d)} Strength of the superconducting order and its projections on the
  honeycomb $d$-wave components as a function of the Hubbard-$U$. The $d$-wave
  order parameter increases exponentially as function of interaction strength
  $U$. The $s^+$ projection has negligible weight.  {\bf{}(e)} Free energy of
  all complex superpositions of the real-valued irreducible representations
  $d_{xy}$ and $d_{x^2-y^2}$. The minima in free energy occur for the
  real-valued nematic solutions (and their $U(1)$ transformations) only.}
  \label{fig:nematic}
\end{figure*}

The only free parameter in our approach is the value of the Hubbard-$U$, which
cannot easily be extracted from first-principles due to the large number of
atoms in the moir\'{e} unit cell. Besides, transport measurements are very
sensitive to the dielectric environment which may screen the interactions more
strongly. This can be achieved by, for example, varying the distance between the
MATTG sample and the metallic gate(s)~\cite{stepanov2020untying,
Goodwin2020screening}, using a dielectric substrate with larger dielectric
constant~\cite{pizarro2019internal} or placing an AB stacked graphene bilayer in
the immediate vicinity of the sample~\cite{liu2021tuning}. In our study, we
adopt $U=5.1\,\mathrm{eV}$, which is a realistic value for graphene-based
materials obtained from first-principles calculations~\cite{Schueler2013,
Wehling2011}.  Choosing this particular value of the Hubbard interaction
supports spin-fluctuation induced superconductivity for $\nu = -2 - \delta$ on
the hole-doped side and $\nu = +2 +\delta $ on the electron-doped side, which is
schematically visualized by the blue superconducting (SC) domes in the top panel
of Fig.~\ref{fig:sttg}~{\bf(e)}. Based on our argument, we propose that the
superconducting instabilities can shift to different fillings if the dielectric
screening reduces the Hubbard-$U$ to smaller values. For example, if the
interaction strength is screened to $U=4.4\,\mathrm{eV}$ as visualized by the
green dashed line in Fig.~\ref{fig:sttg}~{\bf(e)}, superconducting domes shift
towards AFM regions in the phase diagram where $U \lesssim U_{\mathrm{c}}$. As
can be seen in the top panel of the same Figure, this results in the formation
of two superconducting domes (green) at fillings $\nu=-1-\delta$ on the hole
doped side and at $\nu = +1-\delta$ on the electron doped side.

As recent experiments have demonstrated the displacement field tunability of the
superconducting phase~\cite{hao2021engineering, Park2021}, we further
investigate the system in the presence of an electric field
($\Delta_D=\pm30\,\mathrm{meV}$), with the results shown in
Fig.~\ref{fig:sttg}~{\bf (f)}. We observe that the spin-fluctuation spectrum
undergoes major changes such that superconductivity is enhanced at the
electron-doped side and a superconducting dome spreads over almost the entire
filling range between $\nu = +2$ and $\nu = +3$, which is in agreement with
recent experimental findings~\cite{hao2021engineering, Park2021}. This highly
tunable filling dependence of the superconducting domes under the influence of
an electric field is caused by modifications in the low-energy bandstructure of
MATTG~\cite{lopez2020electrical} and hence the spin-fluctuation spectrum. As
depicted in Fig.~\ref{fig:sttg}~{\bf(e,f)}, FM, semi-metallic or metallic states
move to the electron side at $\nu = +2, +3$ and the vacated phase space in
between is taken over by superconducting domes driven by AFM fluctuations in the
paramagnetic phase. We relate these findings to those of the experimental
reports~\cite{hao2021engineering, Park2021} in more detail in the discussion
section below.

\paragraph*{Nematic superconducting order.}---
Next, we analyze the nature of the superconducting spin-singlet order parameter
$\hat{\Delta}$ that is obtained from our atomistic BCS theory for unconventional
superconductivity (see Method section for details). We concentrate on
$U=5.1\,\mathrm{eV}$, $T=0.2\,\mathrm{K}$ and $\nu=-2.5$, firmly placing the
system in a superconducting state. Here, the superconducting order shows clear
nematic signatures of $C_{3z}$-symmetry breaking on the atomic (carbon-carbon)
bond scale and the moir\'{e} length scale, see
Fig.~\ref{fig:nematic}~{\bf(a,b)}. First, the spatially resolved order parameter
amplitude $|\Delta(\bvec{r_i})|$ obtained from averaging the order field
$\hat\Delta = \Delta_{ij}$ over nearest-neighbor bonds, and shown in
Fig.~\ref{fig:nematic}~{\bf(a)}, depicts clear signatures of $C_{3z}$ symmetry
breaking on the moir\'{e} scale.  The nematic superconducting ground-state is
three-fold degenerate consisting of three order parameters with nematic axis
$C_2$ varying by rotation around $120^{\circ}$. All three states are degenerate
in free energy, thus breaking the original $C_{3z}$ symmetry of the Hamiltonian
spontaneously. Furthermore, we find that the order parameter is completely
real-valued and thus restores time-reversal symmetry (TRS) in contrast to any
chiral $d_{x^2-y^2}\pm id_{xy}$ superconducting state. In fact, we find that
complex-linear combinations of the $d$-wave components
$\Delta_{ij}=\cos(\theta)f^{ij}_{d_{x^2-y^2}}+\sin(\theta)e^{i
\phi}f^{ij}_{d_{xy}}$ are energetically disfavored [see
Fig.~\ref{fig:nematic}~{\bf(e)}]. The amplitude distribution of the order
parameter is strongly enhanced in the AAA regions along the nematic $C_2$ axis
in the middle layer of MATTG and is a factor of $\sim10$ smaller in the outer
two layers as depicted in Fig.~\ref{fig:nematic}~{\bf(a)}. At the same time, the
order parameter vanishes in the ABA and BAB regions as expected due to the lack
of states in the non-interacting Hamiltonian. 

In addition, we characterize the superconducting order parameter on the atomic
scale, where the symmetry is given by the $D_{6h}$ point group of the single
graphene sheets. As the FLEX pairing vertex $\hat{\Gamma}_2$ is most attractive
on nearest-neighbor bonds, we project the gap onto the complete basis set
$f_{\eta}$ spanned by the irreducible representations of $D_{6h}$, which
consists of the extended s-wave $\eta = s^+$ and two $d$-wave components
$d_{xy}$ and $d_{x^2-y^2}$, see Methods. Our analysis reveals that the
real-valued order parameter shows nematic $d$-wave characteristics with
vanishing $s$-wave amplitude, similar to atomistic calculations in
MATBG~\cite{fischer2021spin, lothman2021nematic}. To this end, we define a
real-valued two-component vector for each carbon atom $\bvec{\tau}(\bvec r_i) =
\sum_j \Delta_{ij}(f_{d_{x^2-y^2}}^{ij},f_{d_{xy}}^{ij})^{\mathrm{T}}$ that
captures the spatially varying orientation in the $d$-wave components and is
displayed as streamlines in Fig.~\ref{fig:nematic}~{\bf(b)}. There exists a
local $d$-wave nematicity on the carbon-carbon bond scale that forms a
vortex-antivortex structure close to the AAA and ABA/BAB regions and is aligned
to the $C_2$ nematic axis on the moir\'e scale. Interestingly, the phase of
$\bvec\tau_i$ from the middle to the outer layers of MATTG is shifted by $\pi$,
which is a consequence of the interlayer repulsion in the normal-state
Hamiltonian. This $\pi$-locked Josephson coupling was previously also observed
between the two graphene sheets in MATBG and is expected to stabilize the
nematic phase.~\cite{lothman2021nematic, fischer2021spin}.

The averaged amplitude of the nematic superconducting order parameter
$|\bar{\Delta}| = \langle |\Delta(\bvec{r}_i)| \rangle$ is sensitive to the
choice of the Hubbard-$U$ with respect to the critical interaction strength
$U_{\mathrm{c}}$ predicted by the RPA analysis. In
Figure~\ref{fig:nematic}~{\bf(d)}, we show that for $\nu = -2.5$ and $T=0.2 \,
\text{K}$ nematic superconductivity is only present if $U-U_{\mathrm{c}}
\lesssim 0.1 \, \text{eV}$ in our approximation. Approaching the magnetic
instability $U \to U_{\mathrm{c}}$, the overall amplitude of the pairing
interaction $\hat{\Gamma}_2$ increases and the gap parameter grows exponentially
as function of interaction strength and density of states $\rho(\epsilon)$ on
the fermi surface $\propto \exp\big[-1/(\rho(E_F)|\hat{\Gamma}_2| )\big]$ as
expected in a weak-coupling theory. This also emphasizes the sensitivity of
superconductivity to  screening the interactions or to changes in the
spin-fluctuation spectrum, as for example by a displacement field as
demonstrated in Fig.~\ref{fig:sttg}~{\bf(e,f)}.

\begin{figure*}[ht!]
  \includegraphics[width=\textwidth]{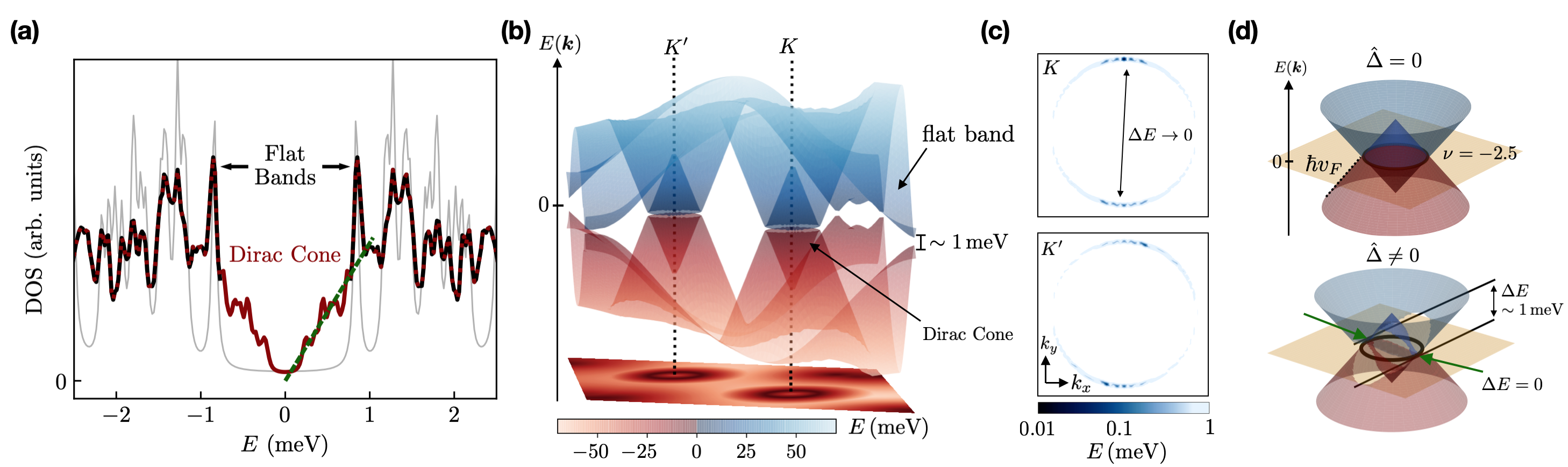}
  \caption{ {\bf{}Quasiparticle density of states (DOS) in the superconducting
  phase of MATTG} for $\nu=-2.5$ and $T= 0.2 \, \text{K}$. {\bf{(a)}} The
  density of states in the superconducting state captures fermionic
  quasiparticle excitations with energies $\pm E_{n, \bvec k} \approx \pm
  \sqrt{(\epsilon_{n, \bvec k}-\mu)^2 + |\Delta_{n, \bvec k}|^2}$ that result in
  the DOS being particle-hole (ph) symmetric. In the DOS of all layers (black
  line), we find clear and separate features of the Dirac cone (highlighted by
  the red line) and the flat bands: While the flat bands are fully-gapped on an
  energy scale of $\sim1$ meV (black line), the Dirac cone leads to separate
  linear signatures $\sim |E|$ (green dahsed line) in the DOS. The partial DOS
  of the middle layer of MATTG (grey line) shows no contribution from the Dirac
  cone's DOS $\sim |E|$, but contains contributions of the flat bands only. This
  is in agreement with the Dirac cone having dominant spectral weight in the
  outer layers, whereas the flat bands dominate in the middle layer.
  {\bf{(b,c)}} Quasiparticle energy landscape $E(\bvec k)$ as function of
  momentum in the Brillouin zone of MATTG. In the superconducting phase, the
  Dirac cones at $K$ and $K^{\prime}$ each show two nodes in the quasiparticle
  spectrum reflecting the $C_{3z}$ symmetry breaking nature of the nematic
  superconducting phase in momentum space. These nodes become distinguishable by
  the dark blue regions in {\bf{(c)}} indicating the vanishing gap amplitude
  $|\Delta_{n, \bvec k}| \to 0$ as the quasiparticle energies approach $E(\bvec
  k)\to 0$. {\bf{(d)}} Schematic sketch of the Dirac cone in the normal-state
  (upper panel) and superconducting phase (lower panel). In the non-interacting
  case $\hat \Delta = 0$ and at filling $\nu=-2.5$, the chemical potential
  (yellow plane) cuts the Dirac cone at energies higher than the Dirac point
  such that the tip of the Dirac cone is mirrored by ph-symmetry and the Fermi
  surface consists of a ring (black line). In the superconducting phase, the
  quasiparticle spectrum shows two nodes at this Fermi surface (green arrows),
  but is gapped away from these points.}
  \label{fig:dos}
\end{figure*}

The nematic properties of the superconducting state lead to clear signatures of
$C_{3z}$-symmetry breaking in the local density of states (LDOS).
Figure~\ref{fig:nematic}~{\bf(c)} depicts the LDOS at the AAA region in the
outer layer of MATTG. We find that the superconducting state gaps the flat bands
in MATTG, while the highly dispersive Dirac cone remains partially ungapped
(within the energy resolution $0.01\,\mathrm{meV}$ of our calculations) as each
Dirac cone shows two nodes on the Fermi surface, see Fig.~\ref{fig:dos}.  For
low energies $E\lesssim |\Delta|$, quasi particle excitations are forbidden in
the AAA regions along the nematic axis $C_2$ due to the large gap amplitude in
the superconducting condensate as shown in the sub panel (2) of Fig.
\ref{fig:nematic}~{\bf (c)}.  For larger energies [sub panels (1), (3), (4)] the
system continuously goes back to a spectral weight distribution similar to the
normal-state Hamiltonian shown in Fig.~\ref{fig:sttg} with a lower degree of
$C_{3z}$ symmetry breaking.  To analyze the behavior of the Dirac cone and the
flat bands in more detail, in Fig.~\ref{fig:dos}~{\bf(a)} we show the density of
states (DOS) in the superconducting phase either for the full system (black
thick line) and restricted to the central layer (thin grey line). In the BCS
formalism, condensation of Cooper pairs occurs on the energy scale of
$|\bar\Delta|$, whereas electronic excitations are described in terms of
fermionic quasiparticles with energy $\pm E_{n, \bvec k} \approx \pm
\sqrt{(\epsilon_{n, \bvec k}-\mu)^2 + |\Delta_{n, \bvec k}|^2}$ that are shifted
relative to the energies in the normal state $\epsilon_{n, \bvec k}$ by the
order parameter amplitude (in momentum-band space) $|\Delta_{n, \bvec k}|$,
resulting in a particle-hole symmetric DOS. In the DOS of all layers, we find
separate features of the Dirac cone (highlighted by the red line) and the flat
bands. While the flat bands are fully-gapped (black line), the Dirac cone
remains partially ungapped and leads to separate linear signatures $\sim |E|$
(red line).

First, the flat bands of MATTG are fully-gapped on an energy scale of
$\sim1\,\mathrm{meV}$ (black line) corresponding to the average gap amplitude
$|\bar\Delta|$. Although the gap can reach values of $10\,\mathrm{meV}$ in the
AAA regions where it is strongly amplified, the DOS is sensitive only to the
spatially averaged value of the gap. Comparing this to the density of states of
only the middle layer (thin grey line) illustrates that the Dirac cone has
dominant weight on the outer layers only: The linear contribution to the density
of states of all layers is not present in the central layer and a true gap of
$\sim1\,\mathrm{meV}$ opens up. This shows that the dominant contribution to the
flat bands resides in the central layer. 

Next, we analyze the band splitting of the Dirac cones by calculating the
quasiparticle spectrum $E(\bvec{k})$ throughout the Brillouin zone of MATTG, see
Fig.~\ref{fig:dos}~{\bf(b)}. The Dirac cones located at $K$ and $K^{\prime}$
each show two nodes in the quasiparticle spectrum that reflect the $C_{3z}$
symmetry breaking of the nematic superconducting state in momentum space. In
Figure~\ref{fig:dos}~{\bf(c)}, these nodes become visible by the dark blue
regions indicating the vanishing gap amplitude $|\bar\Delta| \to 0$ as the
quasiparticle energies approach $E(\bvec k)\to 0$. In
Figure~\ref{fig:dos}~{\bf(d)} we sketch the situation close to the Dirac cone
for the chosen filling of $\nu = -2.5$. The chemical potential (yellow plane)
cuts the Dirac cone at energies higher than the one of the Dirac point such that
the tip of the Dirac Cone is mirrored due to the particle-hole symmetry of the
quasiparticle spectrum and the Fermi surface consists of a ring (black line). In
the superconducting phase, the particle-hole symmetric quasiparticle spectrum
shows two nodes on this Fermi surface [green arrows in panel {\bf(d)}]. Away
from these nodes, the Dirac cone is slightly gapped. Collectively, this leads to
a linearly increasing DOS around the Fermi-level, in contrast to the constant
contributions that would be present if the Dirac Cones remained completely
ungapped.

Since the flat bands (Dirac Cone) can be labelled according to the symmetric
(anti-symmetric) irreducible representation of the mirror reflection symmetry
$\sigma_h$, the different contributions to the DOS suggest that that
superconductivity in MATTG is mainly driven by the flat bands, whereas
superconductivity in the Dirac Cone is only induced by proximity to the flat
bands.  This is also reflected in the real-space properties of the order
parameter $\hat \Delta$.  Since the Dirac Cone belongs to the anti-symmetric
irreducible representation it only has spectral weight on the outer layers
($l=1,3$), where superconductivity is significantly suppressed relative to the
middle layer.  Meanwhile, the superconducting state is fully gapped in the
central layer, which must originate from the flat bands. We hence argue that
quasiparticle excitations in the Dirac Cone are only possible due to proximity
to the flat bands.

\section*{Discussion}
Our work emphasizes the role of spin-fluctuation exchange in the formation of
superconducting instabilities in MATTG. By including both long-ranged (Hartree)
and short-ranged (Hubbard-$U$) electron-electron interactions in our microscopic
theory, we find that without a displacement field, ferromagnetic ordering
dominates around integer fillings $\nu=-3,-2,-1$ and $1$ even at relatively
small interaction strengths ($U\approx 3-4\,\mathrm{eV}$). In between these
integer fillings stabilizing antiferromagnetic order requires a larger
interaction strength ($U\approx 4-5\,\mathrm{eV}$). Intriguingly, estimating $U$
from the monolayer graphene case puts the interaction strength right at the
border of these different ordering tendencies. This allows superconductivity
mediated by spin fluctuations in between integer fillings for values smaller
than the critical interaction required to stabilize AFM order. This
superconductivity is bounded by the above mentioned stronger ferromagnetic order
when approaching integer fillings, naturally restricting the region of
superconductivity to dome structures in a $\nu-T$ phase diagram. 

At zero electrical field, choosing, e.g., $U=5.1\,\mathrm{eV}$ we find the
strongest superconductivity for filling between $\nu=-3$ and $\nu=-2$ caused by
low-energy anti-ferromagnetic spin-fluctuations in the paramagnetic phase. In
this regime, superconductivity occurs within a characteristic dome shape in the
$\nu-T$ phase diagram exhibiting a critical temperature of $T_\mathrm{c} \approx
2\,\mathrm{K}$. This is also where
Refs.~\onlinecite{Park2021,hao2021engineering} find clear signatures of a
superconducting state. A weaker superconducting feature is predicted around
$\nu=+2$, which, however, within our approach is not surrounded by ferromagnetic
order. In Ref.~\onlinecite{Park2021} superconductivity was reported at a similar
filling, while Ref.~\onlinecite{hao2021engineering} only finds very weak
indications of superconductivity around that filling without displacement field. 

Including a displacement field, ferromagnetic order is weakened at the hole-side
and strengthened at the electron side at least for integer fillings of $\nu=+2$
and $+3$, which we attribute to dramatic changes in the spin-fluctuation
spectrum. Superconductivity is strengthened in between these fillings in
agreement with Ref.~\onlinecite{hao2021engineering}. The additional
superconducting features, which appear only in Ref.~\onlinecite{Park2021} at
fillings between $\nu=+1$ and $+2$ and in between $\nu=-1 $ and $-2$ deserve
further study. At values of $U$ for which we can robustly argue for
superconductivity between $\nu=\pm2 $ and $\pm3$, antiferromagnetic order has
already taken over at filling between $\nu=\pm1 $ and $\pm2$. This might hint
towards an insufficiency of our mean-field approach, overestimating the strength
of antiferromagnetic order due to the absence of quantum fluctuations and
neglecting interchannel feedback.  From a methodological point of view, it would
be conducive to study MATTG using unbiased methods like the functional
renormalization group or tensor product states, although this may require
fundamental advance to numerically treat the large moir\'{e} unit cell without
restricting the analysis to the low-energy bands.

As next steps, we suggest the experimental scrutiny of (i) the
filling-dependence of the superconducting phase when tuning the electronic
interactions by screening~\cite{stepanov2020untying} and of (ii) the emergent
nematicity. The latter should yield clear signatures of $C_{3z}$-symmetry
breaking in the LDOS being accessible within STM measurements. Since recent
experimental work~\cite{cao2021large} suggests that the superconducting phase
might be of non-spin-singlet type close to the filling $\nu = -2.4$, this also
deserves further theoretical investigation in our approach as ferromagnetic
spin-fluctuations, which surround the SC dome in our phase diagram, are known to
drive spin-triplet SC phases~\cite{Ran684}. Revealing the intrinsic interplay
between spin-singlet and spin-triplet phases is an exciting avenue of future
research.

\section*{Methods}
\paragraph*{Moir\'e Structure.}--- The commensurate moir\'e unit cell of twisted
trilayer graphene (TTG) can be defined using the same convention as twisted
bilayer graphene (TBG)~\cite{trambly2010localization}. We start from perfectly
aligned, AAA stacked trilayer graphene with a carbon atom of each layer residing
at the origin of the $x-y$ plane, and twist the middle layer anti-clockwise with
respect to the encapsulating layers about the $z$-axis. This creates a structure
with a single moir\'e pattern because of the alignment of the encapsulating
layers. 

For atomistic methods, commensurate moir\'e unit cells must be constructed.
Following Ref.~\citenum{trambly2010localization}, commensurate moir\'e
structures are defined by two integers $n$ and $m$ which specify the twist
angle, $\theta$, via 
\begin{equation}
  \cos\theta=\frac{n^{2}+4 n m+m^{2}}{2\left(n^{2}+n m+m^{2}\right)}.
\end{equation}
The corresponding triangular moi\'e lattice has a length scale that is
determined through the twist angle via
\begin{equation}
  L(\theta)= \dfrac{a_0}{2\sin(\theta/2)},
\end{equation}
where $a_0$ is the lattice constant of graphene. For our simulations, we use
$n,m=20,21$ ($\theta=1.61\degree$) which leads to a commensurate unit cell of
size $L(\theta)=8.59\,\mathrm{nm}$ that contains $N=7566$ carbon sites.

We relax the atomic positions of TTG using classical force fields implemented in
LAMMPS~\cite{LAMMPS}. For the intralayer potential we use
AIREBO-morse~\cite{AIREBO} and for the interlayer potential we use
Kolmogorov–Crespi potential~\cite{KC}. We take the lattice parameter of graphene
to be $a_0 = 2.42~\textrm{\AA}$.

\paragraph*{Atomistic Modelling \& Hartree Calculations.}--- In real space, the
atomistic tight binding Hamiltonian takes the form
\begin{align}
  \begin{split}
    \mathsf H^0 &{}= \sum_{ij,\sigma\sigma} H_{ij}^0c_{i\sigma}^\dagger
      c_{j\sigma}^{\phantom\dagger}, \\
    H_{ij}^0 &{}= t(\bvec r_i - \bvec r_j).
  \end{split}
\end{align}
The operator $c_{i\sigma}^{(\dagger)}$ annihilates (creates) an electron at site
$\bvec r_i$ with spin $\sigma$. The $\mathrm p_z$ electrons are coupled via
Slater-Koster hopping parameters~\cite{trambly2010localization}:
\begin{align}
  \begin{split}
    t(\bvec{d}) &= t_{\parallel}(\bvec{d}) + t_{\bot}(\bvec{d}) \\
    t_{\parallel}(\bvec{d}) &= V_{pp \pi}^0\exp \left(
      -\frac{|\bvec{d}|-a_{\text{cc}}}{\delta_0}\right) \left[1 -
        \left(\frac{\bvec{d} \cdot \bvec{e}_z}{|\bvec{d}|} \right)^2 \right] \\
    t_{\bot}(\bvec{d}) &= V_{pp \sigma}^0\exp \left(
      -\frac{|\bvec{d}|-d_0}{\delta_0}\right)\left[
        \frac{\bvec{d} \cdot \bvec{e}_z}{|\bvec{d}|} \right]^2.
  \end{split}
\label{tb_hop}
\end{align}
Here, $\bvec{e}_z$ is a unit vector which points perpendicular to the graphene
sheets, $d_0 = 1.362 \, a_0$ is the vertical spacing of graphite, $\delta_0 =
0.184 \, a_0$ is the transfer integral decay length, and $a_\mathrm{cc} =
a_0/\sqrt{3}$ is the distance between two nearest neighboring carbon atoms (in
pristine graphene). The term $V_{pp \sigma} =  0.48 \,\mathrm{eV}$ describes the
interlayer hopping while $V_{pp \pi} = - 2.7 \,\mathrm{eV}$ models the
intralayer hopping.

To obtain good agreement with DFT calculations for the electronic structure at
charge neutrality, we add an additional onsite energy term through
\begin{equation}
  H^\Delta_{ii2} = -35\,\mathrm{meV},
\end{equation}
where this term only exists on the inner, twisted layer (2), as indicated by the
subscript of the Hamiltonian.

We performed self-consistent Hartree tight-binding calculations following the
method outlined in Ref.~\citenum{goodwin2020hartree}. Full details of the method
have been outlined in the Supplementary Methods for completeness. We
utilise a 8$\times$8 regular grid in the Brillouin zone to calculate the
electron density and a 11$\times$11 set of moir\'e unit cells to converge the
Hartree potential in real space. An on-site interaction of 17~eV is used with a
$1/|r|$ potential with a dielectric constant of 1 for all other interactions. We
work in the limit of zero temperature and employ a linear mixing scheme to
iteratively converge the set of equations.

The full tight binding Hamiltonian consists of the following contributions
\begin{equation}
  \begin{split}
    \hat H &{}= \hat H^0 + \hat H^H + \hat H^\Delta + \hat H^E, \\
    \mathsf H &{}= \sum_{ij\sigma} H_{ij} c^\dagger_{i\sigma} c^{\phantom{\dagger}}_{j\sigma}.
  \end{split}
\end{equation}
The hopping terms are included through $\hat H^0$, Hartree interactions are
included through $\hat H^H$, the additional onsite potential energy is included
through $\hat H^\Delta$, and finally an electric field is included through $\hat
H^E$. Further details of the Hartree contributions and the inclusion of the
electric field can be found in the Supplementary Methods.

\paragraph*{Density Functional Theory.}--- Density Functional Theory (DFT)
calculations were carried out using \textsc{ONETEP}~\cite{onetep_2020} on large
twist angle TTG structures, see Supplementary Methods for results. We
utilised the Perdew-Burke-Ernzerhof exchange-correlation
functional~\cite{PBE_PRL77} with projector-augmented-wave
pseudopotentials~\cite{GARRITY2014446}, a kinetic energy cutoff of 800~eV, and a
minimal basis of four non-orthogonal generalized Wannier functions per carbon
atom. Due to the metallic nature of TTG, we use the ensemble-DFT
approach~\cite{Serrano_JCP139_2013}. Our calculations are converged such that
the total energy change between iterations is less than $25\,\mathrm{meV}$.

\paragraph*{Magnetic Susceptibility.}--- To account for magnetic fluctuations,
we add a Hubbard interaction acting on the graphene $\mathrm p_z$ orbitals in
the Hamiltonian $\mathsf H$:
\begin{align}
  \mathsf H^\mathrm{U} = \mathsf H + \mathsf H^\mathrm{I}, \quad
  \mathsf H^\mathrm{I} = \sum_{i\sigma} U\,n_{i,\sigma}n_{i,\bar\sigma}.
\end{align}
To treat the four-fermion term $\mathsf H^\mathrm{I}$, we employ the random
phase approximation in the static, long-wavelength limit $q = (\bvec q,
q_0)\to0$ as presented in Ref.~\onlinecite{klebl2019inherited}. To this end, we
calculate the atomistic magnetic susceptibility $\hat\chi_0$:
\begin{equation}
  \hat\chi_0 = \hat\chi_0(\bvec q=q_0=0) = \frac{T}{N_{\bvec k}}
    \sum_{\bvec k,k_0}\hat G(\bvec k,k_0) \hadamard
      \hat G(\bvec k, k_0)^\mathrm{T}.
  \label{eqn:chisum}
\end{equation}
The Green's function $ \hat G(\bvec k,k_0) = G_{ij}(\bvec k,k_0)$ as a function
of Matsubara frequency $k_0$, moir\'e momentum $\bvec k$ is given by
\begin{align}
  \hat{G}(\bvec k,k_0) = (ik_0\mathds1-\hat{H}(\bvec k)+\mu\mathds1)^{-1},
\end{align}
with $\hat H(\bvec k)$ the ``non-interacting" tight-binding Hamiltonian
\emph{including} the Hartree corrections and $\mu$ is the chemical potential
corresponding to the filling factor used in the Hartree potential.

We find that the critical Hubbard-$U$ needed for the onset of magnetic ordering
is given by $U_\mathrm{c}=-1/\lambda_0$, with $\lambda_0$ being the lowest
eigenvalue of $\hat\chi_0$. The magnetic ordering is proportional to the
corresponding eigenvector $\vec v_0$. For numerical evaluation of the Matsubara
sum in Eq.~\eqref{eqn:chisum}, we use the exact same frequency grid presented in
Ref.~\onlinecite{klebl2020importance} and thus are able to take into account the
effect of low-temperature instabilities. We sample the moir\'e Brillouin zone
with $N_{\bvec k} = 24$ points for the RPA simulations.

\paragraph*{Fluctuation-Exchange approximation.}--- To account for pairing
instabilities mediated by charge- and spin-fluctuation exchange, we derive a
microscopic pairing interaction $\hat{\Gamma}_2$ in the fluctuation-exchange
approximation (FLEX) that incorporates effects of transverse and longitudinal
spin-fluctuations. In terms of the full atomistic RPA susceptibility
$\hat\chi_0$, the scattering between Cooper pairs in the singlet channel is
described by the pairing vertex~\cite{Berk1966}
\begin{equation}
  \hat{\Gamma}_2(q) = U \mathds1  - \frac{U^2\hat{\chi}_0(q)}{\mathds1+U
    \hat{\chi}_0(q)}  + \frac{U^3\hat{\chi}_0^2(q)}{\mathds1-U^2
      \hat{\chi}_0^2(q)}.
  \label{flex}
\end{equation}
For diagrammatic details on the derivation of the FLEX pairing vertex, the
reader may refer to the Supplementary Methods. In the static,
long-wavelength limit $q = (\bvec q, q_0)\to0$, the spatial profile of the
long-ranged interaction vertex strongly depends on the magnetic fluctuations
predicted by the RPA analysis, see Supplementary Methods. The latter limit
proves to contain the relevant physics when starting with local repulsive
interactions. The RPA susceptibility $\hat \chi_0$ predicts spin correlations at
length scales intermediate to the carbon-carbon bond scale and moir\'e length
scale, thus being described by orderings at $\bvec q=0$. The system hence shows
the same ordering tendencies in all moir\'{e} unit cells with variable
correlations present on the carbon-carbon bond scale.

\paragraph*{Superconducting state.}--- To analyze the superconducting properties
of the system, we decouple the effective pairing vertex $\hat\Gamma_2$ in
mean-field approximation, allowing only for symmetric spin-singlet bond order
parameters $\Delta_{ij} = \Delta_{ji}$ due to the proximity to AFM tendencies in
between the integer fillings
\begin{multline}
  \hat\Delta(\bvec{k}) = \Delta_{nm}(\bvec{k}) = - \frac{1}{2N}
    \sum_{\bvec{k}^{\prime} \sigma} \, \Gamma_{2,nm}(\bvec{k} -
      \bvec{k}^{\prime}, q_0 \to 0) \\ \times \sigma \langle
        c_{n \sigma}(\bvec{k}^{\prime}) c_{m \bar{\sigma}}(-\bvec{k}^{\prime})
      \rangle_{\text{\tiny MF}}. 
  \label{singlet}
\end{multline}
The resulting mean-field Hamiltonian can be rewritten in the  Nambu spinor basis
$\psi^{\dagger}_{\bvec{k}} = ( \vec{c}^{\phantom \dagger}_{\bvec{k}\uparrow}
\,\,\,  \vec{c}^{\,\dagger}_{-\bvec{k}\downarrow}   )^{\mathrm{T}}.$ The full
$2N$-dimensional Hamiltonian for the spin-dependent $\mathrm{p}_z$ orbitals of
the carbon atoms is of Bogoliubov-de Gennes (BdG) form 
\begin{equation}
  \mathsf{H}_{\text{MF}} = \sum_{\bvec{k}} \psi^{\dagger}_{\bvec{k}}  
    \begin{pmatrix} \hat{H}(\bvec{k}) & \hat{\Delta}(\bvec{k}) \\
    \hat{\Delta}^{\dagger}(\bvec{k}) & -\hat{H}^*(-\bvec{k})  \\\end{pmatrix}
  \psi^{\phantom{\dagger}}_{\bvec{k}} + \text{const.} 
  \label{bdg}
\end{equation}
The BdG bilinear form is diagonalized by a block-structured unitary transform
$\hat{U}_{\bvec{k}}$ 
\begin{equation}
  \begin{split}
  \mathsf{H}_{\text{MF}} &= \sum_{\bvec{k}} \psi^{\dagger}_{\bvec{k}}
    \begin{pmatrix} \hat{H}(\bvec{k}) & \hat{\Delta}(\bvec{k}) \\
      \hat{\Delta}^{\dagger}(\bvec{k}) & -\hat{H}^*(-\bvec{k})  \\\end{pmatrix}
      \psi^{\phantom{\dagger}}_{\bvec{k}} \\ &= \sum_{\bvec{k}} \left (
      \hat{U}_{\bvec{k}} \psi_{\bvec{k}}\right )^{\dagger} \begin{pmatrix}
      \hat{E}_{\bvec{k}} &0 \\ 0 & -\hat{E}_{\bvec{k}}  \end{pmatrix} \left (
      \hat{U}_{\bvec{k}} \psi_{\bvec{k}}\right )^{\phantom \dagger} \\
      \hat{U}^{\phantom \dagger}_{\bvec{k}} &= \begin{pmatrix}
      \hat{u}_{\bvec{k}} & -\hat{v}_{\bvec{k}} \\ \hat{v}^*_{\bvec{k}}&
      \hat{u}_{\bvec{k}}  \\\end{pmatrix} \quad \text{and} \quad
      \hat{U}^{\dagger}_{\bvec{k}} \hat{U}^{\phantom \dagger}_{\bvec{k}} =
      \mathds{1}.
  \end{split}
  \label{bdg_unitary}
\end{equation}
The matrices $ \hat{u}_{\bvec{k}} \left (  \hat{v}_{\bvec{k}} \right )$ are $N
\times N$ matrices, which describe the particle (hole) amplitudes of the
fermionic Bogoliubov quasiparticles $\gamma_{\bvec{k}}$ with energies $\pm
E_{\bvec{k}}$. The latter are defined as
\begin{equation}
  \begin{pmatrix} \gamma^{\phantom \dagger}_{\bvec{k}, \uparrow} \\  \gamma^{
    \dagger}_{-\bvec{k}, \downarrow} \end{pmatrix} = \begin{pmatrix}
    \hat{u}_{\bvec{k}\uparrow} & \hat{v}_{\bvec{k}\uparrow} \\
    -\hat{v}^*_{\bvec{k}\downarrow}& \hat{u}_{\bvec{k}\downarrow}
    \\\end{pmatrix} \begin{pmatrix} c^{\phantom \dagger}_{\bvec{k}, \uparrow} \\
  c^{\dagger}_{-\bvec{k}, \downarrow} \end{pmatrix}.
  \label{bdg_quasi}
\end{equation}
Together with Eq. \eqref{singlet}, this yields a set of non-linear equations
that needs to be solved self-consistently:
\begin{multline}
  \langle c_{i \uparrow}(\bvec k) c_{j \downarrow}(-\bvec k) -  c_{i
  \downarrow}(\bvec k) c_{j \uparrow}(-\bvec k) \rangle_{\text{\tiny MF}}  = \\
  \sum_n \left (u_{ni, \bvec k \uparrow} v_{nj, \bvec k\downarrow}^* + v_{nj,
  \bvec k \uparrow} u_{ni, \bvec k \downarrow}^* \right)\text{tanh}
  \left(\frac{E_{n, \bvec k}}{2T}\right )\,.
\end{multline}
To this end, we start with an initial guess $\Delta_{ij}^{\text{init}}$ and
iterate until convergence is achieved using a linear mixing $\hat\Delta^{n+1} =
(1-\alpha) \hat\Delta^{n} - \alpha \hat\Delta^{n-1}$ scheme to avoid bipartite
solutions in the fixed point iteration. In all of our calculations, we set the
relative error for convergence of $\hat\Delta$ to $\epsilon = 10^{-6}$ and set
the mixing parameter to $\alpha = 0.2$. Here, we only account for the gap
parameter $\hat\Delta$ at the $\Gamma$-point of the Brillouin zone. This
approximation may not change the underlying physics significantly as our
microscopic interaction $\hat\Gamma_2$ in the static, long-wavelength limit ($q
\to 0$) carries no momentum dependence and the size of the Brillouin zone of
MATTG is drastically reduced due to the large real-space unit cell.  Still, we
checked that our results do not change qualitatively when taking a dense mesh
with up to 24 $\bvec{k}$-points in the BZ into account. Furthermore, we track
the free energy $F$ of the system for different initial guesses and during each
self-consistency run to ensure proper convergence of the BdG algorithm, see
Supplementary Methods.

To determine the local amplitude of the superconducting state from the
bond-related order field $\hat\Delta = \Delta_{ij}$, we introduce a
three-dimensional vector $\vec\Delta(\bvec{r}_i) = (\Delta_{i,
i+\bvec{\delta}_1} , \Delta_{i, i+\bvec{\delta}_2}, \Delta_{i,
i+\bvec{\delta}_3})^{\mathrm{T}}$ that contains the superconducting bonds
to all (three) neighboring sites of the carbon atom located at $\bvec{r}_i$. The
norm of this order parameter field yields the lattice-resolved amplitude
$|\Delta(\bvec{r}_i)|$, whereas the overall amplitude $|\bar\Delta| = \langle
|\Delta(\bvec{r}_i)| \rangle$ follows from averaging this quantity over all
sites in the moir\'e unit cell. Here, we only take the $l=1,2,3$ nearest
neighbor carbon-carbon bonds $\bvec{\delta}_l$ into account. This is
equivalent to considering all superconducting bonds $|\bar\Delta| \approx
||\hat\Delta||$ as nearest-neighbor sites are strongly favoured in terms of
attractive interaction generated by the spin-fluctuations exchange mechanism,
see Supplementary Methods.

\paragraph*{Symmetry classification of the gap parameter and LDOS.}--- To
characterize the superconducting order parameter with respect to different
pairing channels on the atomic carbon-carbon bond scale, we project the order
parameter onto the complete basis set formed by the irreducible representations
of the $D_{6h}$ point group
\begin{equation}
  \Delta_{\eta}(n) = \sum_{l} f_{\eta}(\bvec{\delta}_l)(c_{n\uparrow} c_{n
  + \bvec{\delta}_l \downarrow} - c_{n\downarrow} c_{n +
  \bvec{\delta}_l \uparrow}),
  \label{form_factor}
\end{equation}
where $\eta$ denotes different spin-singlet pairing channels: the extended
s-wave $s^{+}$ and two $d$-wave components $d_{xy}$ and $d_{x^2-y^2}$. The
coefficients $f_{\eta}(\bvec{\delta}_l)$ are form factors that correspond
to the pairing channel and are obtained by symmetrizing the bond functions
$\bvec{\delta}_l$ with the irreducible representations of the point group
$D_{6h}$ of graphene. The form factors are: $s^{+} = (1,1,1)/\sqrt{3}$,
$d_{x^2-y^2} = (2,-1,-1)/\sqrt{6}$ and $d_{xy} = (0, 1, -1)/\sqrt{2}$. Here, we
only take the $l=1,2,3$ nearest neighbor carbon-carbon bonds
$\bvec{\delta}_l$ into account, as they are dominant in terms of
attractive interaction strength, see Supplementary Methods. The local
density of states (LDOS) in the superconducting phase is given by
\begin{equation}
  \rho_i(\omega) = \sum_{\bvec{k}, n} |u_{ni,\bvec{k}}|^2 \delta \left
  (\omega-E_{n,\bvec{k}} \right ) + |v_{ni,\bvec{k}}|^2 \delta \left
  (\omega+E_{n,\bvec{k}} \right ).
  \label{ldos}
\end{equation}
To this end, we assume that the gap does not change significantly when
calculated at different points in the (mini)-Brillouin zone (BZ) such that
$\hat\Delta(\bvec{k}) \approx \hat\Delta(\bvec{k} = 0)$. Applying this
$\Gamma$-point approximation, we diagonalize the Nambu Hamiltonian
Eq.~\eqref{bdg} for up to $100 \times 100$ points in the BZ of MATTG. To speed
up the convergence, we use an adaptive momentum mesh that allows for finer
sampling around the $K$ and $K^{\prime}$ points. The $\delta$-function in
Eq.~\eqref{ldos} is approximated by a Lorentzian kernel with broadening $\eta =
0.03$ meV. The density of states (DOS) is obtained from the LDOS by adding up
the contribution of all sites.  To capture the degree of $C_{3z}$ symmetry
breaking in the LDOS in the nematic superconducting phase, we compute the
energy-resolved anisotropy $\zeta^{C_{3z}}(E)$ by comparing the LDOS at each
site $\vec{\rho}(E)$ with its counterpart $R_{\varphi}\rho_{i}(E)$ resulting
from rotation around $\varphi\in\{120^{\circ},240^\circ\}$: \begin{equation}
\zeta(E) = \frac12\sum_{\varphi \in\{2\pi/3,4\pi/3\}}\frac{||
R_{\varphi}\vec{\rho}(E) -  \vec{\rho}(E) ||}{|| \vec{\rho}(E) ||}.
\label{c3_anis} \end{equation} Note that the nematic superconducting order
parameter breaks  $C_{3z}$ symmetry on the microscopic (graphene) sub-lattice
and moir\'{e} scale. Since the latter is easier to access experimentally, we
smear the LDOS in real-space with a Gaussian envelope of width $\sigma = a_0$
such that $\zeta^{C_{3z}}(E)$ only captures the degree of $C_{3z}$ symmetry
breaking on the moir\'{e} scale.

\section*{Data Availability Statement}
Data are available from the corresponding authors upon reasonable request.

\section*{Code Availability Statement}
Simulation Codes are available from the corresponding authors upon reasonable
request.

\section*{Acknowledgements}
We thank V.~Vitale for useful discussions on DFT simulations and X.~Liang for
useful discussions on the relaxations of the system. ZG was supported through a
studentship in the Centre for Doctoral Training on Theory and Simulation of
Materials at Imperial College London funded by the EPSRC (EP/L015579/1). We
acknowledge funding from EPSRC grant EP/S025324/1, support from the Thomas Young
Centre under grant number TYC-101, and the Imperial College London Research
Computing Service (DOI:10.14469/hpc/2232) for computational resources used in
carrying out this work. The Deutsche  Forschungsgemeinschaft (DFG, German
Research Foundation) is acknowledged for support through RTG 1995, within the
Priority Program SPP 2244 “2DMP” and under Germany’s Excellence Strategy-Cluster
of Excellence Matter and Light for Quantum Computing (ML4Q) EXC2004/1 -
390534769. We acknowledge support from the Max Planck-New York City Center for
Non-Equilibrium Quantum Phenomena. RPA, BdG and LDOS calculations were performed
with computing resources granted by RWTH Aachen University under projects
rwth0589 and rwth0595.

\section*{Competing Interests}
The authors declare no competing interests.

\section*{Author Contributions}
A.F., Z.A.H.G. and L.K. performed atomistic modeling. D.M.K. and L.K.
conceptualized and supervised the project. All authors contributed to
interpreting the results and writing the manuscript.

\foreach \x in {1,...,12}
{%
\clearpage
\begin{tikzpicture}[remember picture,overlay,every node/.style={anchor=center}]
  \node at (page cs:0,0) {\includegraphics[page=\x]{supp.pdf}};
\end{tikzpicture}
}


\begin{thebibliography}{10}
\section*{References}
\expandafter\ifx\csname url\endcsname\relax
  \def\url#1{\texttt{#1}}\fi
\expandafter\ifx\csname urlprefix\endcsname\relax\def\urlprefix{URL }\fi
\providecommand{\bibinfo}[2]{#2}
\providecommand{\eprint}[2][]{\url{#2}}

\newcommand\arxiv[2]{%
  \newblock \bibinfo{journal}{\emph{Preprint at} \url{http://arxiv.org/abs/#1}}
    (\bibinfo{year}{#2}).
}

\bibitem{cao2018unconventional}
\bibinfo{author}{Cao, Y.} \emph{et~al.}
\newblock \bibinfo{title}{Unconventional superconductivity in magic-angle
  graphene superlattices}.
\newblock \emph{\bibinfo{journal}{Nature}} \textbf{\bibinfo{volume}{556}},
  \bibinfo{pages}{43--50} (\bibinfo{year}{2018}).

\bibitem{cao2018mott}
\bibinfo{author}{Cao, Y.} \emph{et~al.}
\newblock \bibinfo{title}{Correlated insulator behaviour at half-filling in
  magic-angle graphene superlattices}.
\newblock \emph{\bibinfo{journal}{Nature}} \textbf{\bibinfo{volume}{556}},
  \bibinfo{pages}{80--84} (\bibinfo{year}{2018}).

\bibitem{moiresim}
\bibinfo{author}{Kennes, D.~M.} \emph{et~al.}
\newblock \bibinfo{title}{Moir\'e heterostructures: a condensed matter quantum
  simulator}.
\newblock \emph{\bibinfo{journal}{Nat. Phys.}} \textbf{\bibinfo{volume}{17}},
  \bibinfo{pages}{155–163} (\bibinfo{year}{2021}).

\bibitem{Carr2017twistronics}
\bibinfo{author}{Carr, S.} \emph{et~al.}
\newblock \bibinfo{title}{Twistronics: Manipulating the electronic properties
  of two-dimensional layered structures through their twist angle}.
\newblock \emph{\bibinfo{journal}{Phys. Rev. B}} \textbf{\bibinfo{volume}{95}},
  \bibinfo{pages}{075420} (\bibinfo{year}{2017}).

\bibitem{Guinea2018}
\bibinfo{author}{Guinea, F.} \& \bibinfo{author}{Walet, N.~R.}
\newblock \bibinfo{title}{Electrostatic effects, band distortions, and
  superconductivity in twisted graphene bilayers}.
\newblock \emph{\bibinfo{journal}{Proc. Natl. Acad. Sci. U.S.A.}} \textbf{\bibinfo{volume}{115}},
  \bibinfo{pages}{13174–13179} (\bibinfo{year}{2018}).

\bibitem{Kennes2018}
\bibinfo{author}{Kennes, D.~M.}, \bibinfo{author}{Lischner, J.} \&
  \bibinfo{author}{Karrasch, C.}
\newblock \bibinfo{title}{Strong correlations and d+id superconductivity in
  twisted bilayer graphene}.
\newblock \emph{\bibinfo{journal}{Phys. Rev. B}} \textbf{\bibinfo{volume}{98}},
  \bibinfo{pages}{241407(R)} (\bibinfo{year}{2018}).

\bibitem{fischer2021spin}
\bibinfo{author}{Fischer, A.}, \bibinfo{author}{Klebl, L.},
  \bibinfo{author}{Honerkamp, C.} \& \bibinfo{author}{Kennes, D.~M.}
\newblock \bibinfo{title}{Spin-fluctuation-induced pairing in twisted bilayer
  graphene}.
\newblock \emph{\bibinfo{journal}{Phys. Rev. B}}
  \textbf{\bibinfo{volume}{103}}, \bibinfo{pages}{L041103}
  (\bibinfo{year}{2021}).

\bibitem{klebl2020importance}
\bibinfo{author}{Klebl, L.}, \bibinfo{author}{Goodwin, Z. A.~H.},
  \bibinfo{author}{Mostofi, A.~A.}, \bibinfo{author}{Kennes, D.~M.} \&
  \bibinfo{author}{Lischner, J.}
\newblock \bibinfo{title}{Importance of long-ranged electron-electron
  interactions for the magnetic phase diagram of twisted bilayer graphene}.
\newblock \emph{\bibinfo{journal}{Phys. Rev. B}}
  \textbf{\bibinfo{volume}{103}}, \bibinfo{pages}{195127}
  (\bibinfo{year}{2021}).

\bibitem{klebl2019inherited}
\bibinfo{author}{Klebl, L.} \& \bibinfo{author}{Honerkamp, C.}
\newblock \bibinfo{title}{Inherited and flatband-induced ordering in twisted
  graphene bilayers}.
\newblock \emph{\bibinfo{journal}{Phys. Rev. B}}
  \textbf{\bibinfo{volume}{100}}, \bibinfo{pages}{155145}
  (\bibinfo{year}{2019}).

\bibitem{Gonzalez2017}
\bibinfo{author}{Gonzalez-Arraga, L.~A.}, \bibinfo{author}{Lado, J.~L.},
  \bibinfo{author}{Guinea, F.} \& \bibinfo{author}{San-Jose, P.}
\newblock \bibinfo{title}{Electrically controllable magnetism in twisted
  bilayer graphene}.
\newblock \emph{\bibinfo{journal}{Phys. Rev. Lett.}}
  \textbf{\bibinfo{volume}{119}}, \bibinfo{pages}{107201}
  (\bibinfo{year}{2017}).

\bibitem{Xie2020HF}
\bibinfo{author}{Xie, M.} \& \bibinfo{author}{MacDonald, A.~H.}
\newblock \bibinfo{title}{On the nature of the correlated insulator states in
  twisted bilayer graphene}.
\newblock \emph{\bibinfo{journal}{Phys. Rev. Lett.}}
  \textbf{\bibinfo{volume}{124}}, \bibinfo{pages}{097601}
  (\bibinfo{year}{2020}).

\bibitem{Stauber2019KL}
\bibinfo{author}{Gonz\'alez, J.} \& \bibinfo{author}{Stauber, T.}
\newblock \bibinfo{title}{Kohn-luttinger superconductivity in twisted bilayer
  graphene}.
\newblock \emph{\bibinfo{journal}{Phys. Rev. Lett.}}
  \textbf{\bibinfo{volume}{122}}, \bibinfo{pages}{026801}
  (\bibinfo{year}{2019}).

\bibitem{Zhang2020}
\bibinfo{author}{Zhang, Y.}, \bibinfo{author}{Jiang, K.},
  \bibinfo{author}{Wang, Z.} \& \bibinfo{author}{Zhang, F.}
\newblock \bibinfo{title}{Correlated insulating phases of twisted bilayer
  graphene at commensurate filling fractions: A hartree-fock study}.
\newblock \emph{\bibinfo{journal}{Phys. Rev. B}}
  \textbf{\bibinfo{volume}{102}}, \bibinfo{pages}{035136}
  (\bibinfo{year}{2020}).

\bibitem{Goodwin2019attractive}
\bibinfo{author}{Goodwin, Z. A.~H.}, \bibinfo{author}{Corsetti, F.},
  \bibinfo{author}{Mostofi, A.~A.} \& \bibinfo{author}{Lischner, J.}
\newblock \bibinfo{title}{Attractive electron-electron interactions from
  internal screening in magic angle twisted bilayer graphene}.
\newblock \emph{\bibinfo{journal}{Phys. Rev. B}}
  \textbf{\bibinfo{volume}{100}}, \bibinfo{pages}{235424}
  (\bibinfo{year}{2019}).

\bibitem{Stauber2020}
\bibinfo{author}{Gonz\'alez, J.} \& \bibinfo{author}{Stauber, T.}
\newblock \bibinfo{title}{Time-reversal symmetry breaking versus chiral
  symmetry breaking in twisted bilayer graphene}.
\newblock \emph{\bibinfo{journal}{Phys. Rev. B}}
  \textbf{\bibinfo{volume}{102}}, \bibinfo{pages}{081118(R)}
  (\bibinfo{year}{2020}).

\bibitem{Cea2020}
\bibinfo{author}{Cea, T.} \& \bibinfo{author}{Guinea, F.}
\newblock \bibinfo{title}{Band structure and insulating states driven by
  coulomb interaction in twisted bilayer graphene}.
\newblock \emph{\bibinfo{journal}{Phys. Rev. B}}
  \textbf{\bibinfo{volume}{102}}, \bibinfo{pages}{045107}
  (\bibinfo{year}{2020}).

\bibitem{klebl2020frg}
\bibinfo{author}{Klebl, L.}, \bibinfo{author}{Kennes, D.~M.} \&
  \bibinfo{author}{Honerkamp, C.}
\newblock \bibinfo{title}{Functional renormalization group for a large moir\'e
  unit cell}.
\newblock \emph{\bibinfo{journal}{Phys. Rev. B}}
  \textbf{\bibinfo{volume}{102}}, \bibinfo{pages}{085109}
  (\bibinfo{year}{2020}).

\bibitem{Yan2018}
\bibinfo{author}{Yan, X.}, \bibinfo{author}{Law, X. K.~T.} \&
  \bibinfo{author}{Lee, P.~A.}
\newblock \bibinfo{title}{Kekul\'e valence bond order in an extended hubbard
  model on the honeycomb lattice with possible applications to twisted bilayer
  graphene}.
\newblock \emph{\bibinfo{journal}{Phys. Rev. B}} \textbf{\bibinfo{volume}{98}},
  \bibinfo{pages}{121406(R)} (\bibinfo{year}{2018}).

\bibitem{Wu2018ph}
\bibinfo{author}{Wu, F.}, \bibinfo{author}{MacDonald, A.~H.} \&
  \bibinfo{author}{Martin, I.}
\newblock \bibinfo{title}{Theory of phonon-mediated superconductivity in
  twisted bilayer graphene}.
\newblock \emph{\bibinfo{journal}{Phys. Rev. Lett.}}
  \textbf{\bibinfo{volume}{121}}, \bibinfo{pages}{257001}
  (\bibinfo{year}{2018}).

\bibitem{Choi2018ph}
\bibinfo{author}{Choi, Y.~W.} \& \bibinfo{author}{Choi, H.~J.}
\newblock \bibinfo{title}{Strong electron-phonon coupling, electron-hole
  asymmetry, and nonadiabaticity in magic-angle twisted bilayer graphene}.
\newblock \emph{\bibinfo{journal}{Phys. Rev. B}} \textbf{\bibinfo{volume}{98}},
  \bibinfo{pages}{241412(R)} (\bibinfo{year}{2018}).

\bibitem{Padhi2018WC}
\bibinfo{author}{Padhi, B.}, \bibinfo{author}{Setty, C.} \&
  \bibinfo{author}{Phillips, P.~W.}
\newblock \bibinfo{title}{Doped twisted bilayer graphene near magic angles:
  Proximity to wigner crystallization, not mott insulation}.
\newblock \emph{\bibinfo{journal}{Nano Lett.}} \textbf{\bibinfo{volume}{18}},
  \bibinfo{pages}{6175--6180} (\bibinfo{year}{2018}).

\bibitem{yu2021nematicity}
\bibinfo{author}{Yu, T.}, \bibinfo{author}{Kennes, D.~M.},
  \bibinfo{author}{Rubio, A.} \& \bibinfo{author}{Sentef, M.~A.}
\newblock \bibinfo{title}{Nematicity arising from a chiral superconducting
  ground state in magic-angle twisted bilayer graphene under in-plane magnetic
  fields}. \arxiv{2101.01426}{2021}

\bibitem{samajdar2021electricfieldtunable}
\bibinfo{author}{Samajdar, R.} \emph{et~al.}
\newblock \bibinfo{title}{Electric-field-tunable electronic nematic order in
  twisted double-bilayer graphene}. \arxiv{2102.08385}{2021}

\bibitem{qin2021critical}
\bibinfo{author}{Qin, W.}, \bibinfo{author}{Zou, B.} \&
  \bibinfo{author}{MacDonald, A.~H.}
\newblock \bibinfo{title}{Critical magnetic fields and electron-pairing in
  magic-angle twisted bilayer graphene}. \arxiv{2102.10504}{2021}

\bibitem{lin2019chiral}
\bibinfo{author}{Lin, Y.-P.} \& \bibinfo{author}{Nandkishore, R.~M.}
\newblock \bibinfo{title}{Chiral twist on the high-${T}_{c}$ phase diagram in
  moir\'e heterostructures}.
\newblock \emph{\bibinfo{journal}{Phys. Rev. B}}
  \textbf{\bibinfo{volume}{100}}, \bibinfo{pages}{085136}
  (\bibinfo{year}{2019}).

\bibitem{lin2020parquet}
\bibinfo{author}{Lin, Y.-P.} \& \bibinfo{author}{Nandkishore, R.~M.}
\newblock \bibinfo{title}{Parquet renormalization group analysis of
  weak-coupling instabilities with multiple high-order van hove points inside
  the brillouin zone}.
\newblock \emph{\bibinfo{journal}{Phys. Rev. B}}
  \textbf{\bibinfo{volume}{102}}, \bibinfo{pages}{245122}
  (\bibinfo{year}{2020}).

\bibitem{lu2019superconductors}
\bibinfo{author}{Lu, X.} \emph{et~al.}
\newblock \bibinfo{title}{Superconductors, orbital magnets and correlated
  states in magic-angle bilayer graphene}.
\newblock \emph{\bibinfo{journal}{Nature}} \textbf{\bibinfo{volume}{574}},
  \bibinfo{pages}{653--657} (\bibinfo{year}{2019}).

\bibitem{Cao2020strange}
\bibinfo{author}{Cao, Y.} \emph{et~al.}
\newblock \bibinfo{title}{Strange metal in magic-angle graphene with near
  planckian dissipation}.
\newblock \emph{\bibinfo{journal}{Phys. Rev. Lett.}}
  \textbf{\bibinfo{volume}{124}}, \bibinfo{pages}{076801}
  (\bibinfo{year}{2020}).

\bibitem{Polshyn2019}
\bibinfo{author}{Polshyn, H.} \emph{et~al.}
\newblock \bibinfo{title}{Large linear-in-temperature resistivity in twisted
  bilayer graphene}.
\newblock \emph{\bibinfo{journal}{Nat. Phys.}} \textbf{\bibinfo{volume}{15}},
  \bibinfo{pages}{1011--1016} (\bibinfo{year}{2019}).

\bibitem{yankowitz2019tuning}
\bibinfo{author}{Yankowitz, M.} \emph{et~al.}
\newblock \bibinfo{title}{Tuning superconductivity in twisted bilayer
  graphene}.
\newblock \emph{\bibinfo{journal}{Science}} \textbf{\bibinfo{volume}{363}},
  \bibinfo{pages}{1059--1064} (\bibinfo{year}{2019}).

\bibitem{liu2021tuning}
\bibinfo{author}{Liu, X.} \emph{et~al.}
\newblock \bibinfo{title}{Tuning electron correlation in magic-angle twisted
  bilayer graphene using coulomb screening}.
\newblock \emph{\bibinfo{journal}{Science}} \textbf{\bibinfo{volume}{371}},
  \bibinfo{pages}{1261--1265} (\bibinfo{year}{2021}).

\bibitem{stepanov2020untying}
\bibinfo{author}{Stepanov, P.} \emph{et~al.}
\newblock \bibinfo{title}{Untying the insulating and superconducting orders in
  magic-angle graphene}.
\newblock \emph{\bibinfo{journal}{Nature}} \textbf{\bibinfo{volume}{583}},
  \bibinfo{pages}{375--378} (\bibinfo{year}{2020}).

\bibitem{saito2020independent}
\bibinfo{author}{Saito, Y.}, \bibinfo{author}{Ge, J.},
  \bibinfo{author}{Watanabe, K.}, \bibinfo{author}{Taniguchi, T.} \&
  \bibinfo{author}{Young, A.~F.}
\newblock \bibinfo{title}{Independent superconductors and correlated insulators
  in twisted bilayer graphene}.
\newblock \emph{\bibinfo{journal}{Nature Physics}}
  \textbf{\bibinfo{volume}{16}}, \bibinfo{pages}{926--930}
  (\bibinfo{year}{2020}).

\bibitem{Das2020Chern}
\bibinfo{author}{Das, I.} \emph{et~al.}
\newblock \bibinfo{title}{Symmetry-broken chern insulators and rashba-like
  landau-level crossings in magic-angle bilayer graphene}.
\newblock \emph{\bibinfo{journal}{Nature Physics}}
  \textbf{\bibinfo{volume}{17}}, \bibinfo{pages}{710--714}
  (\bibinfo{year}{2021}).

\bibitem{Wu2020Chern}
\bibinfo{author}{Wu, S.}, \bibinfo{author}{Zhang, Z.},
  \bibinfo{author}{Watanabe, K.}, \bibinfo{author}{Taniguchi, T.} \&
  \bibinfo{author}{Andrei, E.~Y.}
\newblock \bibinfo{title}{Chern insulators, van hove singularities and
  topological flat bands in magic-angle twisted bilayer graphene}.
\newblock \emph{\bibinfo{journal}{Nature materials}}
  \textbf{\bibinfo{volume}{20}}, \bibinfo{pages}{488--494}
  (\bibinfo{year}{2021}).

\bibitem{Nuckolls2020Chern}
\bibinfo{author}{Nuckolls, K.~P.} \emph{et~al.}
\newblock \bibinfo{title}{Strongly correlated chern insulators in magic-angle
  twisted bilayer graphene}.
\newblock \emph{\bibinfo{journal}{Nature}} \textbf{\bibinfo{volume}{588}},
  \bibinfo{pages}{610--615} (\bibinfo{year}{2020}).

\bibitem{arora2020superconductivity}
\bibinfo{author}{Arora, H.~S.} \emph{et~al.}
\newblock \bibinfo{title}{Superconductivity in metallic twisted bilayer
  graphene stabilized by wse 2}.
\newblock \emph{\bibinfo{journal}{Nature}} \textbf{\bibinfo{volume}{583}},
  \bibinfo{pages}{379--384} (\bibinfo{year}{2020}).

\bibitem{Serlin2020}
\bibinfo{author}{Serlin, M.} \emph{et~al.}
\newblock \bibinfo{title}{Intrinsic quantized anomalous hall effect in a
  moir\'e heterostructure}.
\newblock \emph{\bibinfo{journal}{Science}} \textbf{\bibinfo{volume}{367}},
  \bibinfo{pages}{900–903} (\bibinfo{year}{2020}).

\bibitem{Sharpe2019ferro}
\bibinfo{author}{Sharpe, A.~L.} \emph{et~al.}
\newblock \bibinfo{title}{Emergent ferromagnetism near three-quarters filling
  in twisted bilayer graphene}.
\newblock \emph{\bibinfo{journal}{Science}} \textbf{\bibinfo{volume}{365}},
  \bibinfo{pages}{605–608} (\bibinfo{year}{2019}).

\bibitem{Zondiner2020}
\bibinfo{author}{Zondiner, U.} \emph{et~al.}
\newblock \bibinfo{title}{Cascade of phase transitions and dirac revivals in
  magic-angle graphene}.
\newblock \emph{\bibinfo{journal}{Nature}} \textbf{\bibinfo{volume}{582}},
  \bibinfo{pages}{203--208} (\bibinfo{year}{2020}).

\bibitem{Wong2020}
\bibinfo{author}{Wong, D.} \emph{et~al.}
\newblock \bibinfo{title}{Cascade of electronic transitions in magic-angle
  twisted bilayer graphene}.
\newblock \emph{\bibinfo{journal}{Nature}} \textbf{\bibinfo{volume}{582}},
  \bibinfo{pages}{198–202} (\bibinfo{year}{2020}).

\bibitem{Xie2019spectrosopic}
\bibinfo{author}{Xie, Y.} \emph{et~al.}
\newblock \bibinfo{title}{Spectroscopic signatures of many-body correlations in
  magic-angle twisted bilayer graphene}.
\newblock \emph{\bibinfo{journal}{Nature}} \textbf{\bibinfo{volume}{572}},
  \bibinfo{pages}{101--105} (\bibinfo{year}{2019}).

\bibitem{Kerelsky2019maximized}
\bibinfo{author}{Kerelsky, A.} \emph{et~al.}
\newblock \bibinfo{title}{Maximized electron interactions at the magic angle in
  twisted bilayer graphene}.
\newblock \emph{\bibinfo{journal}{Nature}} \textbf{\bibinfo{volume}{572}},
  \bibinfo{pages}{95--100} (\bibinfo{year}{2019}).

\bibitem{Jiang2019charge}
\bibinfo{author}{Jiang, Y.} \emph{et~al.}
\newblock \bibinfo{title}{Charge order and broken rotational symmetry in
  magic-angle twisted bilayer graphene}.
\newblock \emph{\bibinfo{journal}{Nature}} \textbf{\bibinfo{volume}{573}},
  \bibinfo{pages}{91--95} (\bibinfo{year}{2019}).

\bibitem{Choi2019correlations}
\bibinfo{author}{Choi, Y.} \emph{et~al.}
\newblock \bibinfo{title}{Electronic correlations in twisted bilayer graphene
  near the magic angle}.
\newblock \emph{\bibinfo{journal}{Nat. Phys.}} \textbf{\bibinfo{volume}{15}},
  \bibinfo{pages}{1174--1180} (\bibinfo{year}{2019}).

\bibitem{cao2020nematicity}
\bibinfo{author}{Cao, Y.} \emph{et~al.}
\newblock \bibinfo{title}{Nematicity and competing orders in superconducting
  magic-angle graphene}.
\newblock \emph{\bibinfo{journal}{science}} \textbf{\bibinfo{volume}{372}},
  \bibinfo{pages}{264--271} (\bibinfo{year}{2021}).

\bibitem{Carr2020NatRev}
\bibinfo{author}{Carr, S.}, \bibinfo{author}{Fang, S.} \&
  \bibinfo{author}{Kaxiras, E.}
\newblock \bibinfo{title}{Electronic-structure methods for twisted moir\'e
  layers}.
\newblock \emph{\bibinfo{journal}{Nat. Rev. Mater}}
  \textbf{\bibinfo{volume}{5}}, \bibinfo{pages}{748--763}
  (\bibinfo{year}{2020}).

\bibitem{Balents2020rev}
\bibinfo{author}{Balents, L.}, \bibinfo{author}{Dean, C.~R.},
  \bibinfo{author}{Efetov, D.~K.} \& \bibinfo{author}{Young, A.~F.}
\newblock \bibinfo{title}{Superconductivity and strong correlations in moir\'e
  flat bands}.
\newblock \emph{\bibinfo{journal}{Nat. Phys.}} \textbf{\bibinfo{volume}{16}},
  \bibinfo{pages}{725–733} (\bibinfo{year}{2020}).

\bibitem{burg2019correlated}
\bibinfo{author}{Burg, G.~W.} \emph{et~al.}
\newblock \bibinfo{title}{Correlated insulating states in twisted double
  bilayer graphene}.
\newblock \emph{\bibinfo{journal}{Phys. Rev. Lett.}}
  \textbf{\bibinfo{volume}{123}}, \bibinfo{pages}{197702}
  (\bibinfo{year}{2019}).

\bibitem{shen2020correlated}
\bibinfo{author}{Shen, C.} \emph{et~al.}
\newblock \bibinfo{title}{Correlated states in twisted double bilayer
  graphene}.
\newblock \emph{\bibinfo{journal}{Nature Physics}}
  \textbf{\bibinfo{volume}{16}}, \bibinfo{pages}{520--525}
  (\bibinfo{year}{2020}).

\bibitem{liu2020tunable}
\bibinfo{author}{Liu, X.} \emph{et~al.}
\newblock \bibinfo{title}{Tunable spin-polarized correlated states in twisted
  double bilayer graphene}.
\newblock \emph{\bibinfo{journal}{Nature}} \textbf{\bibinfo{volume}{583}},
  \bibinfo{pages}{221--225} (\bibinfo{year}{2020}).

\bibitem{rubioverdu2020universal}
\bibinfo{author}{Rubio-Verdú, C.} \emph{et~al.}
\newblock \bibinfo{title}{Universal moir\'e nematic phase in twisted graphitic
  systems} (\bibinfo{year}{2020}).
\newblock \eprint{2009.11645}.

\bibitem{Park2021}
\bibinfo{author}{Park, J.~M.}, \bibinfo{author}{Cao, Y.},
  \bibinfo{author}{Watanabe, K.}, \bibinfo{author}{Taniguchi, T.} \&
  \bibinfo{author}{Jarillo-Herrero, P.}
\newblock \bibinfo{title}{Tunable strongly coupled superconductivity in
  magic-angle twisted trilayer graphene}.
\newblock \emph{\bibinfo{journal}{Nature}} \textbf{\bibinfo{volume}{590}},
  \bibinfo{pages}{249--255} (\bibinfo{year}{2021}).

\bibitem{hao2021engineering}
\bibinfo{author}{Hao, Z.} \emph{et~al.}
\newblock \bibinfo{title}{Electric field--tunable superconductivity in
  alternating-twist magic-angle trilayer graphene}.
\newblock \emph{\bibinfo{journal}{Science}} \textbf{\bibinfo{volume}{371}},
  \bibinfo{pages}{1133--1138} (\bibinfo{year}{2021}).

\bibitem{cao2021large}
\bibinfo{author}{Cao, Y.}, \bibinfo{author}{Park, J.~M.},
  \bibinfo{author}{Watanabe, K.}, \bibinfo{author}{Taniguchi, T.} \&
  \bibinfo{author}{Jarillo-Herrero, P.}
\newblock \bibinfo{title}{Pauli-limit violation and re-entrant
  superconductivity in moir{\'e} graphene}.
\newblock \emph{\bibinfo{journal}{Nature}} \textbf{\bibinfo{volume}{595}},
  \bibinfo{pages}{526--531} (\bibinfo{year}{2021}).

\bibitem{tsai2019correlated}
\bibinfo{author}{Zhang, X.} \emph{et~al.}
\newblock \bibinfo{title}{Correlated insulating states and transport signature
  of superconductivity in twisted trilayer graphene superlattices}.
\newblock \emph{\bibinfo{journal}{Phys. Rev. Lett.}}
  \textbf{\bibinfo{volume}{127}}, \bibinfo{pages}{166802}
  (\bibinfo{year}{2021}).

\bibitem{shi2020tunable}
\bibinfo{author}{Shi, Y.} \emph{et~al.}
\newblock \bibinfo{title}{Tunable van hove singularities and correlated states
  in twisted trilayer graphene}.
\newblock \emph{\bibinfo{journal}{Nature Physics}}  (\bibinfo{year}{2020}).

\bibitem{wang2020correlated}
\bibinfo{author}{Wang, L.} \emph{et~al.}
\newblock \bibinfo{title}{Correlated electronic phases in twisted bilayer
  transition metal dichalcogenides}.
\newblock \emph{\bibinfo{journal}{Nature materials}} \bibinfo{pages}{1--6}
  (\bibinfo{year}{2020}).

\bibitem{tang2020}
\bibinfo{author}{Tang, Y.} \emph{et~al.}
\newblock \bibinfo{title}{Simulation of hubbard model physics in wse$_2$/ws$_2$
  moir\'e superlattices}.
\newblock \emph{\bibinfo{journal}{Nature}} \textbf{\bibinfo{volume}{579}},
  \bibinfo{pages}{353--358} (\bibinfo{year}{2020}).

\bibitem{xian20}
\bibinfo{author}{Xian, L.} \emph{et~al.}
\newblock \bibinfo{title}{Realization of nearly dispersionless bands with
  strong orbital anisotropy from destructive interference in twisted bilayer
  mos2}.
\newblock \emph{\bibinfo{journal}{Nature communications}}
  \textbf{\bibinfo{volume}{12}}, \bibinfo{pages}{1--9} (\bibinfo{year}{2021}).

\bibitem{regan2020}
\bibinfo{author}{Regan, E.~C.} \emph{et~al.}
\newblock \bibinfo{title}{Mott and generalized wigner crystal states in
  wse$_2$/ws$_2$ moir\'e superlattices}.
\newblock \emph{\bibinfo{journal}{Nature}} \textbf{\bibinfo{volume}{579}},
  \bibinfo{pages}{359--363} (\bibinfo{year}{2020}).

\bibitem{Vitale2021}
\bibinfo{author}{Vitale, V.}, \bibinfo{author}{Atalar, K.},
  \bibinfo{author}{Mostofi, A.~A.} \& \bibinfo{author}{Lischner, J.}
\newblock \bibinfo{title}{Flat band properties of twisted transition metal
  dichalcogenide homo- and heterobilayers of mos2, mose2, ws2 and wse2}.
  \arxiv{2102.03259}{2021}

\bibitem{Xian18}
\bibinfo{author}{Xian, L.}, \bibinfo{author}{Kennes, D.~M.},
  \bibinfo{author}{Tancogne-Dejean, N.}, \bibinfo{author}{Altarelli, M.} \&
  \bibinfo{author}{Rubio, A.}
\newblock \bibinfo{title}{Multiflat bands and strong correlations in twisted
  bilayer boron nitride: Doping-induced correlated insulator and
  superconductor}.
\newblock \emph{\bibinfo{journal}{Nano Letters}} \textbf{\bibinfo{volume}{19}},
  \bibinfo{pages}{4934--4940} (\bibinfo{year}{2019}).

\bibitem{Ni2019}
\bibinfo{author}{Ni, G.~X.} \emph{et~al.}
\newblock \bibinfo{title}{Soliton superlattices in twisted hexagonal boron
  nitride}.
\newblock \emph{\bibinfo{journal}{Nature Communications}}
  \textbf{\bibinfo{volume}{10}}, \bibinfo{pages}{4360} (\bibinfo{year}{2019}).

\bibitem{kennes19}
\bibinfo{author}{Kennes, D.~M.}, \bibinfo{author}{Xian, L.},
  \bibinfo{author}{Claassen, M.} \& \bibinfo{author}{Rubio, A.}
\newblock \bibinfo{title}{One-dimensional flat bands in twisted bilayer
  germanium selenide}.
\newblock \emph{\bibinfo{journal}{Nature Communications}}
  \textbf{\bibinfo{volume}{11}}, \bibinfo{pages}{1124} (\bibinfo{year}{2020}).

\bibitem{Carr2018Pressure}
\bibinfo{author}{Carr, S.}, \bibinfo{author}{Fang, S.},
  \bibinfo{author}{Jarillo-Herrero, P.} \& \bibinfo{author}{Kaxiras, E.}
\newblock \bibinfo{title}{Pressure dependence of the magic twist angle in
  graphene superlattices}.
\newblock \emph{\bibinfo{journal}{Phys. Rev. B}} \textbf{\bibinfo{volume}{98}},
  \bibinfo{pages}{085144} (\bibinfo{year}{2018}).

\bibitem{Goodwin2020screening}
\bibinfo{author}{Goodwin, Z. A.~H.} \emph{et~al.}
\newblock \bibinfo{title}{Critical role of device geometry for the phase
  diagram of twisted bilayer graphene}.
\newblock \emph{\bibinfo{journal}{Phys. Rev. B}}
  \textbf{\bibinfo{volume}{101}}, \bibinfo{pages}{165110}
  (\bibinfo{year}{2020}).

\bibitem{bistritzer2011moire}
\bibinfo{author}{Bistritzer, R.} \& \bibinfo{author}{MacDonald, A.~H.}
\newblock \bibinfo{title}{Moir{\'e} bands in twisted double-layer graphene}.
\newblock \emph{\bibinfo{journal}{Proc. Natl. Acad. Sci. U.S.A.}} \textbf{\bibinfo{volume}{108}}, \bibinfo{pages}{12233--12237}
  (\bibinfo{year}{2011}).

\bibitem{dos2007graphene}
\bibinfo{author}{Dos~Santos, J.~L.}, \bibinfo{author}{Peres, N.} \&
  \bibinfo{author}{Neto, A.~C.}
\newblock \bibinfo{title}{Graphene bilayer with a twist: Electronic structure}.
\newblock \emph{\bibinfo{journal}{Phys. Rev. Lett.}}
  \textbf{\bibinfo{volume}{99}}, \bibinfo{pages}{256802}
  (\bibinfo{year}{2007}).

\bibitem{shallcross2010electronic}
\bibinfo{author}{Shallcross, S.}, \bibinfo{author}{Sharma, S.},
  \bibinfo{author}{Kandelaki, E.} \& \bibinfo{author}{Pankratov, O.}
\newblock \bibinfo{title}{Electronic structure of turbostratic graphene}.
\newblock \emph{\bibinfo{journal}{Phys. Rev. B}}
  \textbf{\bibinfo{volume}{81}}, \bibinfo{pages}{165105}
  (\bibinfo{year}{2010}).

\bibitem{morell2010flat}
\bibinfo{author}{Morell, E.~S.}, \bibinfo{author}{Correa, J.},
  \bibinfo{author}{Vargas, P.}, \bibinfo{author}{Pacheco, M.} \&
  \bibinfo{author}{Barticevic, Z.}
\newblock \bibinfo{title}{Flat bands in slightly twisted bilayer graphene:
  Tight-binding calculations}.
\newblock \emph{\bibinfo{journal}{Phys. Rev. B}}
  \textbf{\bibinfo{volume}{82}}, \bibinfo{pages}{121407}
  (\bibinfo{year}{2010}).

\bibitem{Kang2019strong}
\bibinfo{author}{Kang, J.} \& \bibinfo{author}{Vafek, O.}
\newblock \bibinfo{title}{Strong coupling phases of partially filled twisted
  bilayer graphene narrow bands}.
\newblock \emph{\bibinfo{journal}{Phys. Rev. Lett.}}
  \textbf{\bibinfo{volume}{122}}, \bibinfo{pages}{246401}
  (\bibinfo{year}{2019}).

\bibitem{park2020flavour}
\bibinfo{author}{Park, J.~M.}, \bibinfo{author}{Cao, Y.},
  \bibinfo{author}{Watanabe, K.}, \bibinfo{author}{Taniguchi, T.} \&
  \bibinfo{author}{Jarillo-Herrero, P.}
\newblock \bibinfo{title}{Flavour hund's coupling, correlated chern gaps, and
  diffusivity in moir\'e flat bands}.
\newblock \emph{\bibinfo{journal}{arXiv preprint arXiv:2008.12296}}
  (\bibinfo{year}{2020}).

\bibitem{Chen2019ABC}
\bibinfo{author}{Chen, G.} \emph{et~al.}
\newblock \bibinfo{title}{Signatures of tunable superconductivity in a trilayer
  graphene moir\'e superlattice}.
\newblock \emph{\bibinfo{journal}{Nature}} \textbf{\bibinfo{volume}{572}},
  \bibinfo{pages}{215--219} (\bibinfo{year}{2019}).

\bibitem{Kruchkov2020}
\bibinfo{author}{Carr, S.} \emph{et~al.}
\newblock \bibinfo{title}{Ultraheavy and ultrarelativistic dirac quasiparticles
  in sandwiched graphenes}.
\newblock \emph{\bibinfo{journal}{Nano Lett.}} \textbf{\bibinfo{volume}{20}},
  \bibinfo{pages}{3030--3038} (\bibinfo{year}{2020}).

\bibitem{lopez2020electrical}
\bibinfo{author}{Lopez-Bezanilla, A.} \& \bibinfo{author}{Lado, J.~L.}
\newblock \bibinfo{title}{Electrical band flattening, valley flux, and
  superconductivity in twisted trilayer graphene}.
\newblock \emph{\bibinfo{journal}{Phys. Rev. Research}}
  \textbf{\bibinfo{volume}{2}}, \bibinfo{pages}{033357} (\bibinfo{year}{2020}).

\bibitem{trambly2010localization}
\bibinfo{author}{Trambly~de Laissardi{\`e}re, G.}, \bibinfo{author}{Mayou, D.}
  \& \bibinfo{author}{Magaud, L.}
\newblock \bibinfo{title}{Localization of dirac electrons in rotated graphene
  bilayers}.
\newblock \emph{\bibinfo{journal}{Nano letters}} \textbf{\bibinfo{volume}{10}},
  \bibinfo{pages}{804--808} (\bibinfo{year}{2010}).

\bibitem{zhu2020twisted}
\bibinfo{author}{Zhu, Z.}, \bibinfo{author}{Carr, S.},
  \bibinfo{author}{Massatt, D.}, \bibinfo{author}{Luskin, M.} \&
  \bibinfo{author}{Kaxiras, E.}
\newblock \bibinfo{title}{Twisted trilayer graphene: A precisely tunable
  platform for correlated electrons}.
\newblock \emph{\bibinfo{journal}{Phys. Rev. Lett.}}
  \textbf{\bibinfo{volume}{125}}, \bibinfo{pages}{116404}
  (\bibinfo{year}{2020}).

\bibitem{khalaf2019magic}
\bibinfo{author}{Khalaf, E.}, \bibinfo{author}{Kruchkov, A.~J.},
  \bibinfo{author}{Tarnopolsky, G.} \& \bibinfo{author}{Vishwanath, A.}
\newblock \bibinfo{title}{Magic angle hierarchy in twisted graphene
  multilayers}.
\newblock \emph{\bibinfo{journal}{Phys. Rev. B}}
  \textbf{\bibinfo{volume}{100}}, \bibinfo{pages}{085109}
  (\bibinfo{year}{2019}).

\bibitem{goodwin2020hartree}
\bibinfo{author}{Goodwin, Z.~A.}, \bibinfo{author}{Vitale, V.},
  \bibinfo{author}{Liang, X.}, \bibinfo{author}{Mostofi, A.~A.} \&
  \bibinfo{author}{Lischner, J.}
\newblock \bibinfo{title}{Hartree theory calculations of quasiparticle
  properties in twisted bilayer graphene}.
\newblock \emph{\bibinfo{journal}{Electronic Structure}}
  \textbf{\bibinfo{volume}{2}}, \bibinfo{pages}{034001} (\bibinfo{year}{2020}).

\bibitem{Berk1966}
\bibinfo{author}{Berk, N.~F.} \& \bibinfo{author}{Schrieffer, J.~R.}
\newblock \bibinfo{title}{Effect of ferromagnetic spin correlations on
  superconductivity}.
\newblock \emph{\bibinfo{journal}{Phys. Rev. Lett.}}
  \textbf{\bibinfo{volume}{17}}, \bibinfo{pages}{433--435}
  (\bibinfo{year}{1966}).

\bibitem{pizarro2019internal}
\bibinfo{author}{Pizarro, J.~M.}, \bibinfo{author}{R\"osner, M.},
  \bibinfo{author}{Thomale, R.}, \bibinfo{author}{Valent\'{\i}, R.} \&
  \bibinfo{author}{Wehling, T.~O.}
\newblock \bibinfo{title}{Internal screening and dielectric engineering in
  magic-angle twisted bilayer graphene}.
\newblock \emph{\bibinfo{journal}{Phys. Rev. B}}
  \textbf{\bibinfo{volume}{100}}, \bibinfo{pages}{161102}
  (\bibinfo{year}{2019}).

\bibitem{Schueler2013}
\bibinfo{author}{Sch\"uler, M.}, \bibinfo{author}{R\"osner, M.},
  \bibinfo{author}{Wehling, T.~O.}, \bibinfo{author}{Lichtenstein, A.~I.} \&
  \bibinfo{author}{Katsnelson, M.~I.}
\newblock \bibinfo{title}{Optimal hubbard models for materials with nonlocal
  coulomb interactions: Graphene, silicene, and benzene}.
\newblock \emph{\bibinfo{journal}{Phys. Rev. Lett.}}
  \textbf{\bibinfo{volume}{111}}, \bibinfo{pages}{036601}
  (\bibinfo{year}{2013}).

\bibitem{Wehling2011}
\bibinfo{author}{Wehling, T.~O.} \emph{et~al.}
\newblock \bibinfo{title}{Strength of effective coulomb interactions in
  graphene and graphite}.
\newblock \emph{\bibinfo{journal}{Phys. Rev. Lett.}}
  \textbf{\bibinfo{volume}{106}}, \bibinfo{pages}{236805}
  (\bibinfo{year}{2011}).

\bibitem{lothman2021nematic}
\bibinfo{author}{L{\"o}thman, T.}, \bibinfo{author}{Schmidt, J.},
  \bibinfo{author}{Parhizgar, F.} \& \bibinfo{author}{Black-Schaffer, A.~M.}
\newblock \bibinfo{title}{Nematic superconductivity in magic-angle twisted
  bilayer graphene from atomistic modeling}.
  \arxiv{2101.11555}{2021}

\bibitem{Ran684}
\bibinfo{author}{Ran, S.} \emph{et~al.}
\newblock \bibinfo{title}{Nearly ferromagnetic spin-triplet superconductivity}.
\newblock \emph{\bibinfo{journal}{Science}} \textbf{\bibinfo{volume}{365}},
  \bibinfo{pages}{684--687} (\bibinfo{year}{2019}).

\bibitem{LAMMPS}
\bibinfo{author}{Plimpton, S.}
\newblock \bibinfo{title}{Fast parallel algorithms for short-range molecular
  dynamics}.
\newblock \emph{\bibinfo{journal}{J. Comp. Phys.}}
  \textbf{\bibinfo{volume}{117}}, \bibinfo{pages}{1--19}
  (\bibinfo{year}{1995}).

\bibitem{AIREBO}
\bibinfo{author}{O’Connor, T.~C.}, \bibinfo{author}{Andzelm, J.} \&
  \bibinfo{author}{Robbins, M.~O.}
\newblock \bibinfo{title}{Airebo-m: A reactive model for hydrocarbons at
  extreme pressures}.
\newblock \emph{\bibinfo{journal}{J. Chem. Phys.}}
  \textbf{\bibinfo{volume}{142}}, \bibinfo{pages}{024903}
  (\bibinfo{year}{2015}).

\bibitem{KC}
\bibinfo{author}{Kolmogorov, A.~N.} \& \bibinfo{author}{Crespi, V.~H.}
\newblock \bibinfo{title}{Registry-dependent interlayer potential for graphitic
  systems}.
\newblock \emph{\bibinfo{journal}{Phys. Rev. B}} \textbf{\bibinfo{volume}{71}},
  \bibinfo{pages}{235415} (\bibinfo{year}{2005}).

\bibitem{onetep_2020}
\bibinfo{author}{Prentice, J. C. A.~t.}
\newblock \bibinfo{title}{The onetep linear-scaling density functional theory
  program}.
\newblock \emph{\bibinfo{journal}{J. Chem. Phys.}}
  \textbf{\bibinfo{volume}{152}}, \bibinfo{pages}{174111}
  (\bibinfo{year}{2020}).

\bibitem{PBE_PRL77}
\bibinfo{author}{Perdew, J.~P.}, \bibinfo{author}{Burke, K.} \&
  \bibinfo{author}{Ernzerhof, M.}
\newblock \bibinfo{title}{Generalized gradient approximation made simple}.
\newblock \emph{\bibinfo{journal}{Phys. Rev. Lett.}}
  \textbf{\bibinfo{volume}{77}}, \bibinfo{pages}{3865--3868}
  (\bibinfo{year}{1996}).

\bibitem{GARRITY2014446}
\bibinfo{author}{Garrity, K.~F.}, \bibinfo{author}{Bennett, J.~W.},
  \bibinfo{author}{Rabe, K.~M.} \& \bibinfo{author}{Vanderbilt, D.}
\newblock \bibinfo{title}{Pseudopotentials for high-throughput dft
  calculations}.
\newblock \emph{\bibinfo{journal}{Computational Materials Science}}
  \textbf{\bibinfo{volume}{81}}, \bibinfo{pages}{446--452}
  (\bibinfo{year}{2014}).

\bibitem{Serrano_JCP139_2013}
\bibinfo{author}{Ruiz-Serrano, {\`A}.} \& \bibinfo{author}{Skylaris, C.-K.}
\newblock \bibinfo{title}{A variational method for density functional theory
  calculations on metallic systems with thousands of atoms}.
\newblock \emph{\bibinfo{journal}{J. Chem. Phys.}}
  \textbf{\bibinfo{volume}{139}}, \bibinfo{pages}{054107}
  (\bibinfo{year}{2013}).

\end{thebibliography}
\end{document}